\documentclass[12pt]{article}
\usepackage{amsmath}
\usepackage{graphicx,psfrag,epsf}
\usepackage{natbib}
\usepackage[shortlabels]{enumitem}

\usepackage{float}
\usepackage{amsmath,amssymb,amsthm}
\setcitestyle{authoryear}
\usepackage{booktabs}
\usepackage{bbm}
\usepackage{multicol}
\usepackage[ruled,vlined]{algorithm2e}
\SetKwInOut{Data}{$\triangleright$\ Data}
\SetKwInOut{Input}{$\triangleright$\ Input}
\SetKwInOut{Output}{$\triangleright$\ Output}
\SetKwComment{Comment}{$\triangleright$\ }{}

\SetCommentSty{comsty}

\let\tilde\widetilde
\let\hat\widehat
\def\Exp{{\ensuremath{\mathbb E}}}

\usepackage{helvet}
\usepackage{newtxtext,newtxmath}
\usepackage[scaled]{beramono}
\usepackage{stfloats}
\usepackage[labelfont=bf,tableposition=top]{caption}
\usepackage{titlesec}
\titleformat{\section}{\normalfont\normalsize\bfseries\sffamily}{\thesection.}{.5em}{}
\titleformat{\subsection}{\normalfont\normalsize\itshape\sffamily}{\thesubsection.}{.5em}{}
\titleformat{\subsubsection}{\normalfont\normalsize\itshape\sffamily}{\thesubsubsection.}{.5em}{}
\titlespacing{\section}{0pt}{*4}{*1}
\titlespacing{\subsection}{0pt}{*4}{*1}
\titlespacing{\subsubsection}{0pt}{*4}{*1}

\newcommand{\argmax}{\operatornamewithlimits{argmax}}
\usepackage{threeparttable}
\usepackage[colorlinks,linkcolor=red, citecolor=blue]{hyperref}
\usepackage{enumitem}

\usepackage{stackengine,graphicx}
\stackMath
\newlength\matfield
\newlength\tmplength

\newlist{Properties}{enumerate}{2}
\setlist[Properties]{label=Property \arabic*., font=\textbf, itemindent=*}

\newcommand{\blind}{0}

\begin{document}

\def\spacingset#1{\renewcommand{\baselinestretch}%
{#1}\small\normalsize} \spacingset{1}


\date{}
\if0\blind
{
  \title{\bf Computational Approaches for Exponential-Family Factor Analysis}
  \author{Liang Wang\\
    Email: \texttt{leonwang@bu.edu}\\
    Department of Mathematics and Statistics, Boston University\\
    and \\
    Luis Carvalho \\
    Email: \texttt{lecarval@bu.edu} \\
    Department of Mathematics and Statistics, Boston University}
  \maketitle
} \fi

\if1\blind
{
  \bigskip
  \bigskip
  \bigskip
  \begin{center}
    {\LARGE\bf Title}
\end{center}
  \medskip
} \fi

\bigskip
\begin{abstract}
We study a general factor analysis framework where the $n$-by-$p$ data matrix
is assumed to follow a general exponential family distribution entry-wise.
While this model framework has been proposed before, we here further relax its
distributional assumption by using a quasi-likelihood setup. By parameterizing
the mean-variance relationship on data entries, we additionally introduce a
dispersion parameter and entry-wise weights to model large variations and
missing values. The resulting model is thus not only robust to distribution
misspecification but also more flexible and able to capture mean-dependent
covariance structures of the data matrix.
Our main focus is on efficient computational approaches to perform the factor
analysis. Previous modeling frameworks rely on simulated maximum likelihood
(SML) to find the factorization solution, but this method was shown to
lead to asymptotic bias when the simulated sample size grows slower than
the square root of the sample size $n$, eliminating its practical
application for data matrices with large $n$.
Borrowing from expectation-maximization (EM) and stochastic gradient descent
(SGD), we investigate three estimation procedures based on iterative
factorization updates. Our proposed solution does not show asymptotic biases,
and scales even better for large matrix factorizations with error $O(1/p)$.
To support our findings, we conduct simulation experiments and discuss its
application in four case studies.
\end{abstract}

\noindent%

\spacingset{1.45}

\section{Introduction}
\label{sec:intro}
Over the past decades, factor analysis has gained tremendous attention
in psychology~\citep{1986psyfactor}, computer science~\citep{2008facefactor},
finance~\citep{2015famma}, and biological research~\citep{2021biofactor}.
In particular, when the data $X \in \mathbb{R}^{n\times p}$ is high dimensional
($n \ll p$), effectively modeling and estimating the covariance structure has
been problematic~\citep{1994factor}. The factor model provides an effective
approach to model high dimensional data in which the covariance of the
observations is assumed to lie on a lower dimensional manifold.

Despite its popularity in modeling high dimensional data, factor models
have several limitations. First and foremost, both data and latent
variables are assumed to follow a Gaussian distribution, which is not ideal
for modeling binary, count, or other non-constant variance data. To address
the first limitation, there exist some prior
works that extend the factor model with more general exponential family
assumption~\citep{2001expfactor2,2003expfactor}. However, even with improved
assumptions away from Gaussianity, exponential family distributions
are often too restrictive for real world, overly dispersed data.
Moreover, as we show later in Section~\ref{subsubsec:sml_problem}, the
proposed maximum likelihood estimation algorithm for such an extended model is
problematic with both numerical and asymptotic convergence issues.
As a minor issue, the latent factors are only identifiable up to a rotational
transformation, potentially causing problems in interpreting the latent
factors. Lastly, both these extended works and the traditionally factor
analysis framework lack the flexibility to model missing data, preventing
several interesting applications such as matrix completion.

This paper thus aims at extending previous works by:
\begin{itemize}
  \item Proposing a flexible factor analysis model that assumes only an
  entry-wise robust mean-variance relationship (as in quasi-likelihood
  methods), column-wise dispersion parameters, and entry-wise weights;

  \item Providing interpretability for latent factors via orthogonal
  identifiability constraints, improving the algorithmic robustness of factor
  estimation;

  \item Discussing efficient computational approaches that balance
  computational complexity when the number of factors is (i) small, using
  approximate expectation-maximization, and (ii) large, using stochastic
  gradient descent with adaptive momentum estimation (Adam).
\end{itemize}

To introduce appropriate notations and to understand some of the relevant
attempts to address those issues, we elaborate below on the limitations of
factor models along with some existing remedies proposed in the literature
that motivated our generalization.

\subsection{The Factor Model and Its Limitations}
For given data $X \in \mathbb{R}^{n\times p}$, the traditional rank $q$ factor
model assumes that $q \ll p$ and that the data is generated by
latent factors $\Lambda \in \mathbb{R}^{n\times q}$ with
$\Lambda^\top = [\Lambda_1 \cdots \Lambda_n]$, and a deterministic projection
matrix $V \in \mathbb{R}^{p\times q}$, the loading matrix.
We implicitly assume the following data generating process: for each
observation $i = 1, \ldots, n$:
\begin{align}
\label{eq:factmodel}
\begin{split}
\Lambda_i & \stackrel{\text{iid}}{\sim} N(0, I_q), \\
X_i \,|\, \Lambda_i & \stackrel{\text{ind}}{\sim} N(V \Lambda_i, \Phi)
\end{split}
\end{align}
where $\Phi$ is a $p$-th order symmetric positive definite matrix,
a covariance matrix providing potential heterogeneous noise.

Marginally, $X_i \stackrel{\text{ind}}{\sim} N(0, \Phi + V V^\top)$, so
maximum likelihood estimation of $V$ and $\Phi$ is equivalent to covariance
estimation. It is common to assume that $\Phi$ is diagonal so as to not
confound the effect of the loadings $V$~\citep{bartholomew2011}, and so we
adopt the same assumption from now on. In this case, the MLE estimator for $V$
can be obtained in closed form using matrix calculus, based on the
eigen-decomposition of the data covariance. Alternatively, $V$ and $\Phi$ can
be estimated via expectation-maximization, especially if some of the entries
in the data matrix $X$ are missing.

The model in general has interesting connections to matrix factorization. For
example, probabilistic PCA~\citep{1998ppca} can be considered as equivalent to
the normal factor model with the only difference that the factor model permits
heterogeneous noise structure through the specification of $\Phi =
\text{Diag}\{\psi_1, \ldots, \psi_p\}$. Under an isotropic noise structure,
$\Phi = \psi I_p$, \citet{1963factor} established the connection between these
two models by demonstrating that the stationary point solution of the
factor model likelihood spans the columns of the sample covariance
eigenvectors. Drawing further the analogy from the relationship between
probabilistic PCA and PCA, the factor model can be considered as the random
counterpart of matrix factorization by allowing the factorized components (or
latent local factors) $\Lambda$ to be random.

While finding a wide range of applications, the factor model and its
deterministic counterpart (matrix factorization) have, however, some
limitations.
We discuss them below, including a brief summary on some recent improvements,
along with our proposed solutions to further generalize the factor model.

\subsubsection{Relaxing the restrictive distributional assumption}
\label{subsubsec:relax_gaussian}
In a factor model setup, both the latent variable $\Lambda_i$
and the data are assumed to (conditionally) follow a Gaussian distribution,
yielding a marginal Gaussian distribution for the data. Both assumptions
require careful examination when dealing with real data.

For the data distribution, assuming simply a Gaussian data likelihood
overlooks many interesting structures in the data. For example, network
adjacency matrices take only binary values of 0 and 1, while computer images
take a integer values for pixel intensities. Both types of data have
been shown to be better modeled with discrete distributions from the
exponential family~\citep{2021dmf}. One obvious relaxation is thus to extend
the \emph{data} likelihood assumption from Gaussian to exponential families,
or, to accommodate more robust specifications, to specify mean and covariance
structures, as in quasi-likelihood approaches~\citep[][but check also
Section~\ref{subsec:likeli} for details]{wedderburn1974quasi}.
Moreover, these exponential family generalizations do not consider the
flexible covariance modeling of the data matrix, which has shown to be one of
the most important applications of the Gaussian factor
model~\citep{fan2008high}.

The latent variable assumption is usually considered less restrictive when
compared to the likelihood assumption, as evidenced from similar Gaussian
latent structures in hierarchical statistical models (e.g. the random effects
model~\citep{borenstein2010basic} and the state space
model~\citep{carter1996markov}). This assumption is however frequently
studied together with factor
identifiability~\citep{shapiro1985identifiability} to ensure unique latent
representations of the data. Specifically, the factor
model~\eqref{eq:factmodel} is not identifiable (or unique) since for any
orthogonal matrix $T \in O(q)$,
$\Lambda^*_i \doteq T \Lambda_i \stackrel{\text{iid}}{\sim} N(0, I_q)$ and,
with $V^* = VT^\top$, $X_i \,|\, \Lambda^*_i \stackrel{\text{ind}}{\sim}
N(V^* \Lambda^*_i, \Phi)$ specify the same model since $V\Lambda_i = VT^\top T
\Lambda_i = V^* \Lambda^*_i$, that is, $\Lambda^* = \Lambda T^\top$ and $V^*$
are not identifiable from $\Lambda$ and $V$. For this reason it is common in
factor analysis to rotate factors after fitting the model to achieve better
sparsity and/or interpretability, e.g. with varimax
rotation~\citep{kaiser1958varimax,rohe2023vintage}. However, it is
advantageous to address these identifiability issues from the outset to reduce
the space of potential solutions and speed up estimation procedures. In this
case, it is helpful to borrow from the matrix factorization research. For
example, adding various factorization constraints such as
sparsity~\citep{gribonval2010dictionary}, positivity~\citep{1999nmf} and
orthogonality~\citep{li2010nonnegative} was shown to provide more
representative and unique latent factors. The stochastic counterpart of these
factor constraints is closely related to an evolving research field related to
data manifolds~\citep{ma2012manifold}.

\subsubsection{Allowing entry-wise weight and link transformation}
\label{subsubsec:entry_weight}
Another potential improvement that has remained absent from factor analysis
research is the specification of entry-wise likelihood weights and non-linear
transformations. In matrix factorization, allowing entry-wise factorization
weights and the flexibility of transforming the original data has been shown to
be valuable in providing more representative factorized results. For example,
in the field of natural language processing, Global Vectors for Word
Representation~\citep[GloVe;][]{2014glove} received great success in obtaining
word embeddings. The method essentially applied a log transformation on
the word-occurrence matrix with heuristic entry-wise weights to avoid
over and underweighting toward rare and common word co-occurrences.
In the field of computer vision~\citep{kalayeh2014nmf}, weighting matrices
related to classification class frequencies are introduced to alleviate issues
with class imbalance. This residual boosting weight matrix provides latent
factors that are more suitable for downstream classification. In the field of
matrix completion~\citep{davenport20141}, specification of zero factorization
weights can eliminate missing entries from the factorization, which in turn
allows the latent structure to impute them.

Perhaps due to the computational complexity associated with these
enhancements, such flexibility has not been transferred from matrix
factorization to factor analysis. Link transformations might appear in
the literature, e.g.~\citep{reimann2002factor}, but are mostly applied
as an ad-hoc pre-processing methodology. In practice, it is clear that
entry-wise factor weights and link transformations could greatly
improve the flexibility of the factor modeling framework. Nevertheless, a
unified factor modeling framework that enables such flexibility is still
missing from the literature.

\subsubsection{Improving on optimization procedures for large datasets}
\label{subsubsec:sml_problem}
Lastly, as we seek to improve on the traditional Gaussian factor model,
it is natural to consider practical computational concerns: can we scale
fitting the improved model to modern large datasets?

While the marginal likelihood under model~\eqref{eq:factmodel} is available in
closed form, deriving the marginal likelihood under a non-Gaussian data
assumption is typically difficult and recent research have resorted to
simulated maximum likelihood~\citep[SML;][]{2001expfactor2}, Markov chain
Monte Carlo (MCMC) or variational inference~\citep{2015hpf, computers13060151}.
However, these methods have their own difficulties. Variational inference is
based on an approximation to the target marginal distribution and usually
relies on oversimplified representations for computational gains at the cost
of poor representativity. As for MCMC, due to the identifiability issue
introduced earlier, the marginal likelihood is constant along high
dimensional quotient spaces on $V$ imposed by equivalence under orthogonal
operations (rotations). These equivalent spaces cause challenges
for both the MCMC sampling and the assessment of convergence.
Lastly, although theoretically attractive, MCMC is computationally
expensive since it usually requires long running times to achieve convergence
up to a desired precision when compared to other approaches such as Laplace
approximations~\citep{2009inla}.

As the original optimization method proposed with the initial exponential
factor generalization~\citep{2001expfactor2,2003adfactor}, the SML approach
is considered as one of the most common estimation methods. Specifically, the
maximum likelihood estimator is obtained by maximizing the following
simulated likelihood based on $S$ Monte Carlo samples,
\[
L(V; X) = \prod_{i=1}^n f_V(X_i) \approx
\prod_{i=1}^n \frac{1}{S} \sum_{s=1}^S f_V\big(X_i \,|\, \Lambda_i^{(s)}\big)
\doteq L_{\text{MC}}(V; X),
\]
where $\Lambda_i^{(s)} \stackrel{\text{iid}}{\sim} N(0, I_q)$. To obtain the
maximizer of $\log L_{\text{MC}}(V; X)$, the gradient is needed (up to a
constant):
\begin{equation}
\nabla_{V} \sum_{i=1}^n \log \sum_{s=1}^S f_V(X_i \,|\, \Lambda_i^{(s)}) =
\sum_{i=1}^n \frac{\sum_{s=1}^S \nabla_{V} f_V(X_i \,|\, \Lambda_i^{(s)})}
{\sum_{s=1}^S f_V(X_i \,|\, \Lambda_i^{(s)})}.
\label{eq:sml_grad}
\end{equation}
Despite the fact that $ \nabla_{V} f_V(X_i|\Lambda_i^{(s)})$ is readily known
in closed form, optimization using~\eqref{eq:sml_grad} has both numerical
and theoretical issues. For the numerical issue, we need to observe that the
likelihood $f_V(X_i |\Lambda_i^{(s)})$ evaluations in the denominator need to
be performed in log space to avoid underflows and usually require good
starting points for $V$, which are particularly challenging when the data
dimension $p$ is large.

From a theoretical perspective, the likelihood along its gradient evaluation
depends heavily on the asymptotic behavior of sample size $S$. It has been
shown in~\citep{1995smlsize} that the estimator will be asymptotically biased
if the MC sample size $S$ does not grow faster than data sample size
$\sqrt{n}$. In fact, we verified with numerical studies that the gradient
estimation can potentially require a larger MC sample size $S \gg \sqrt{n}$ to
stabilize $\nabla_{V} f_\theta(X_i|\Lambda_i^{(s)})$ in~\eqref{eq:sml_grad}.
Consequently, when optimizing the likelihood via SML, there is a trade-off
between computation efficiency and estimation bias.  For modern applications
of large data dimensions, the sample size $S$ required to control the
MC variance can be quite large, thus preventing the practical applications of
such methods.

\subsubsection{Real world applications}
Although PCA has been traditionally used for many real world applications \citep{li2024comprehensive}, in the past decades, the generalization of the deterministic PCA factorization to the exponential family has enabled a
series of benchmark models across different fields. For example, the non-negative matrix factorization~\citep[NMF;][]{1999nmf} in computer vision generalized the data
distributional assumption to Poisson. The non-Gaussian state space
model~\citep{kitagawa1987non} in time series generalized the data
distributional assumption to non-Gaussian using non-parametric estimation; the
Skip-gram model~\citep{levy2014neural, mikolov2013efficient} in natural
language processing generalized the data assumption to multinomial. Perhaps
most relevant to statistics factor model research, \citep{2001expfactor2}
and~\citep{2003adfactor} generalized the data likelihood to exponential family
distribution while allowing the random specification of a latent factor~$\Lambda$.

Perhaps due to the infeasibility of the SML estimation described in the
previous subsection, the lack of a practical estimation method has limited the
application of the random factorization to only Bernoulli factor models with
an identity link function, i.e, the random dot product
model~\citep[RDPM;][]{hoff2002latent, young2007random}. Despite its
restrictive identity link assumption, the RDPM has established its
popularity on its empirical evidence from network analysis. After addressing
the SML estimation problem, we also feel that empirical evidence of such a
generalized model has still been missing from the literature.

\subsection{Organization of the paper}
The paper is organized as follows: in Section~\ref{sec:setup} we introduce a
more general \emph{exponential} factor model that addresses the shortcomings
listed in~\ref{subsubsec:relax_gaussian} and~\ref{subsubsec:entry_weight};
next, in Section~\ref{sec:optim}, we discuss our main contributions---a
collection of efficient and robust optimization strategies for inference,
tackling the points in~\ref{subsubsec:sml_problem};
Section~\ref{sec:example} demonstrates the effectiveness of our factorization
with simulated examples and applications on benchmark data from various
fields; finally, Section~\ref{sec:conclusion} concludes with a summary of the
innovations and directions for future work.
\section{Exponential Factor Models}
\label{sec:setup}
\subsection{Guaranteeing factor identifiability}
\label{subsec:identi}
We start by addressing the model issues raised in Section~\ref{sec:intro}.
Given our concern with computational efficiency, our first issue is
non-identifiability; as discussed in~\ref{subsubsec:relax_gaussian}, we need
to constraint the factors to avoid lack of identifiability due to rotations.
From now on we adopt the following standardization of the factors:

\begin{enumerate}
\item[(i)] $\Lambda_i \stackrel{\text{iid}}{\sim} N(0, I_q)$
for $i \in [n]$ as usual, with the distribution of the rows of
$\Lambda$ being invariant to orthogonal transformations;

\item[(ii)] $V$ has scaled pairwise orthogonal columns, that is, $V= U D$
with $U \in \mathcal{S}_{p, q}(\mathbb{R})$, a $p$-frame in the Stiefel
manifold of order $q$, and $D = \text{Diag}_{j \in [q]}\{d_j\}$ with
$d_1 \geq \cdots \geq d_q >0$, so that $V^\top V = D^2$. We denote this
space for $V$ as $\tilde{\mathcal{S}}_{p, q}(\mathbb{R})$.
\end{enumerate}

This setup makes the factorization model identifiable since for any arbitrary
$T \in O(q)$, $V^* = V T^\top$ can only belong to
$\tilde{\mathcal{S}}_{p,q}(\mathbb{R})$ if $T^\top$ commutes with a diagonal
matrix, that is, if $T \in O(1)^q$, and so $V$ is unique (up to column sign
changes, as in the SVD). In practice, given arbitrary loadings $\hat{V}$ we
just need to find the singular value decomposition of $\hat{V}^\top = W D U^\top$
and set $W = I_q$ to identify $\hat{V} = U D$.
Conversely, we can define arbitrary loadings $\hat{V}$ by post-multiplying
$UD$ by an orthogonal matrix $W^\top$; for example, $W$ can be defined as a
varimax rotation to achieve sparsity and thus better interpretability.
For a more in-depth discussion about suitable rotations,
see~\citep{bartholomew2011}.

\subsection{Generalizing the normal likelihood}
\label{subsec:likeli}
Next, we relax the convenient but often unrealistic Gaussian assumptions in
the likelihood and settle with a more general mean and variance specification
in the spirit of \emph{quasi-likelihood}~\citep{wedderburn1974quasi}. We assume that
$X_i \,|\, \Lambda_i \sim F(g^{-1}(V \Lambda_i + \eta_0), \Phi_i)$ where $F$ belongs to
the exponential family with link function $g$ and variance function $\mathcal{V}$,
that is,
\begin{equation}
\label{eq:generate}
    \begin{aligned}
    \Exp(X_i | \Lambda_i) & \doteq \mu_i = g^{-1}(\eta_i),
      \text{~with~} \eta_i = V \Lambda_i + \eta_{0}, \quad\text{and} \\
    \text{Var}(X_i | \Lambda_i) & = \Phi_i \mathbb{V}(\mu_i),\\
    \end{aligned}
\end{equation}
where $\mathbb{V}(\mu_i) = \text{Diag}\{\mathcal{V}(\mu_i)\}$ is the diagonal
variance and $\eta_{0} \in \mathbb{R}^p$ is the latent center of the factor
model.
This way, we can more naturally represent data $X$ belonging to fields other
than real numbers; common cases are binary data with $F$ being Bernoulli or
binomial (with weights) and count data with $F$ being Poisson or negative
binomial. In particular, the negative binomial distribution offers
enhancements over the Poisson distribution by effectively accommodating the
over-dispersion characteristic often observed in count data; see, e.g.,
\citep{xia2020average} for a detailed treatment of negative binomial
distributions as a compound Poisson type.

To accommodate entry-wise weights, as motivated in
Section~\ref{subsubsec:entry_weight}, we set $\Phi_i = \Phi W_i^{-1}$ where
$W_i = \text{Diag}_{j=1,\ldots,p}\{w_{ij}\}$ are the known weights, that is,
$\Phi_i = \text{Diag}_{j=1,\ldots,p}\{\phi_j / w_{ij}\}$.
This setup implies $\Exp(X_{ij}|\Lambda_i) = \mu_{ij}$ and
$\text{Var}(X_{ij} | \Lambda_i) = \phi_j\mathcal{V}(\mu_{ij}) / w_{ij}$.
From these two moment conditions, we can adopt the extended
quasi-likelihood~\citep{nelder1987extended} to define:
\begin{equation}
\label{eq:quasi_family}
    \log f_{V, \eta_0, \Phi}(X_i|\Lambda_i) =
    -\sum_{j=1}^p \frac{w_{ij}}{\phi_j} \int_{\mu_{ij}}^{X_{ij}} \frac{X_{ij} - t}{\mathcal{V}(t)} dt -
    \frac{1}{2}\log\bigg(2\pi \frac{\phi_j\mathcal{V}(X_{ij})}{w_{ij}} \bigg).
\end{equation}
We call this the exponential factor model (\textbf{EFM}).

The MLE estimate of $\theta = (V, \eta_0, \Phi)$ then requires access to the marginal
density for each observation $i \in [n]$,
\begin{equation}
    \log f_{\theta}(X_i) = \int_{\Lambda_i}  \log f_{\theta}(X_i|\Lambda_i) f(\Lambda_i) d\Lambda_i
\end{equation}
where $f$ is the density of the standard multivariate normal density of
order $q$:
\begin{equation}\label{eq:opti}
    \hat{\theta} = \argmax_{V \in \tilde{\mathcal{S}}_{p,q}(\mathbb{R}), \eta_0 \in \mathbb{R}^p,
    \phi_1, \ldots, \phi_p > 0} \sum_{i=1}^n \log f_{\theta}(X_{i}).
\end{equation}

\subsection{Modeling covariance}
Such a generalized factor framework can be used to efficiently estimate the
covariance structure of high dimensional data. Specifically, applying the
total variance formula we can derive the covariance of a new observation
$X \,|\, \lambda \sim F(V\lambda, \Phi)$ given $\lambda \sim N(0, I_q)$,
\begin{equation}
\label{eq:total_cov}
\text{Cov}_{\theta}(X) =
\Exp_{\lambda} \big(\text{Var}_{\theta}(X|\lambda) \big) +
\text{Var}_{\lambda}\big( \Exp_{\theta}(X| \lambda) \big).
\end{equation}
Here, the first term is a diagonal matrix but the second term requires an
outer product that induces correlations among the entries of $X$,
\begin{equation}
\begin{aligned}
\Exp_{\lambda} \big( \text{Var}_{\theta} (X | \lambda) \big) &=
\text{Diag}_{j \in [p]} \Big\{ \phi_j \Exp_{\lambda} \big[\mathcal{V} \circ g^{-1}
\big((V\lambda)_j\big) \big] \Big\}, \\
\text{Var}_{\lambda}\big( \Exp_{\theta}(X| \lambda) \big) &=
\Exp_{\lambda} \Big[\big(g^{-1}(V\lambda) - \mu_\lambda\big)
\big(g^{-1}(V\lambda) - \mu_\lambda\big)^\top \Big],
\end{aligned}
\label{eq:quasi_cov}
\end{equation}
where $\mu_\lambda = \Exp_{\lambda}\big(g^{-1}(V\lambda)\big)$.

In the special case of the Gaussian distribution, we have $g(\mu) = \mu$ and $\mathcal{V}(\mu) = 1$
and so, as expected, $\text{Cov}_{\theta}(X) = \Phi + V V^\top$.
Using this result, \citet{fan2008high} demonstrated the efficiency of the
plug-in covariance estimator via estimation of $V, \phi, \text{Cov}(\Lambda)$.
Similarly, we demonstrate empirically in Section~\ref{ssec:covariance} that
the plug-in efficiency of $(V, \Phi)$ in high dimensional covariance
estimation is preserved under our generalized quasi-factor setup using
Eq~\eqref{eq:quasi_cov}.

\section{Approximate Optimization}
\label{sec:optim}
As we mentioned in Section~\ref{subsubsec:sml_problem}, directly optimizing
Eq~\eqref{eq:opti} through simulated gradients has both numerical and
theoretical issues. Next, we demonstrate how we can conduct maximum likelihood
estimation on the factor matrix $V$ and conditional variances $\Phi$ through
some approximate but efficient and robust algorithms.

\subsection{EM optimization with small rank}
\label{ssec:em}
As in the Gaussian factor model, one common estimation procedure for such a
latent space model should be
Expectation-Maximization~\citep[EM;][]{dempster1977em}. The EM algorithm is
simpler to implement since it depends only on the joint
$f_\theta(X, \Lambda)$ and enjoys higher numerical stability when compared to
gradient-based optimization routines. However, while each EM step is lighter
computationally, the method might converge to multiple modes or suffer from
slower convergence rates~(see, e.g.~\citep{mclachlan2008algorithm} for an
extensive treatment).

In our setup, the EM algorithm can be formulated as follows:

\begin{description}
  \item[E-Step:]
  Given the parameters $\theta^{(t)}$ at the $t$-th iteration step, we compute
  \begin{equation}
  \label{eq:emopti_estep}
  Q(\theta; \theta^{(t)}) = \Exp_{\Lambda|X; \theta^{(t)}} \Bigg[
  \sum_{i=1}^n \log f_{\theta}(X_i, \Lambda_i) \Bigg] 
  \approx \sum_{i=1}^n \int \log f_{\theta}(X_i, \Lambda_i)
  \tilde{f}_{\theta^{(t)}}(\Lambda_i | X_i) d\Lambda_i,
  \end{equation}
  where $\tilde{f}_{\theta^{(t)}}(\Lambda_i | X_i)$ is a Laplace
  approximation~\citep{1986accurate} to the actual conditional posterior on
  $\Lambda_i | X_i$ using a Gaussian density centered at $\hat{\Lambda}_i$ and
  with precision matrix $\hat{H}_i$. These parameters are found by Fisher
  scoring on a regularized GLM regression with quasi-likelihood
  \begin{equation}
  \label{eq:em_elli}
  \ell_i^{(t)}(\Lambda_i; X_i, \theta^{(t)}) =
  \log f_{\theta^{(t)}} (X_i | \Lambda_i) + \log f(\Lambda_i),
  \end{equation}
  that is,
  $\Lambda_i | X_i \approx N\big(\hat{\Lambda}_i(X_i, \theta^{(t)}),\,
  \hat{H}_i^{-1}(X_i, \theta^{(t)})\big)$ with
  \[
  \hat{\Lambda}_i(X_i, \theta^{(t)}) =
  \argmax_{\Lambda_i} \ell_i^{(t)} (\Lambda_i; X_i, \theta^{(t)})
  \quad \text{and} \quad
  \hat{H}_i(X_i, \theta^{(t)}) =
  \Exp_{X_i}\Bigg[-\frac{\partial^2
      \ell_i^{(t)}(\hat{\Lambda}_i; X_i, \theta^{(t)})}%
      {\partial \Lambda_i \partial \Lambda_i^\top}\Bigg]
  \]
  the negative expected Hessian; details can be found in the Appendix. Here we
  make the dependency on data $X_i$ and current parameters $\theta^{(t)}$
  explicit, but, from now on, to avoid overloading the notation we will
  simplify the posterior mode and precision to $\hat{\Lambda}_i$ an
  $\hat{H}_i$.

  When the rank $q$ is small we can evaluate the integral numerically
  in~\eqref{eq:emopti_estep} using multivariate Gauss-Hermite
  cubature~\citep{golub1969gauss,moustaki2000generalized}: with $m$ cubature
  nodes $\Lambda_{il}$ for each $\Lambda_i$,
  \begin{equation}\label{eq:emopti_gq}
  Q(\theta; \theta^{(t)}) \approx
  \sum_{i=1}^n \sum_{l=1}^m \log f_{\theta}(X_i, \Lambda_{il}) w_{il}^{(t)},
  \end{equation}
  where the cubature weights $w_{il}^{(t)}$ are computed based on the Laplace
  approximation $\Lambda_i | X_i; \theta^{(t)} \approx N(\hat{\Lambda}_i,
  \hat{H}_i^{-1})$.

  \item[M-Step:]
  We can then update the parameters to $\theta^{(t+1)}$ by maximizing the
  expected complete data likelihood,
  \begin{equation}\label{eq:emopti}
  \theta^{(t+1)} = \argmax_{\theta = (V, \Phi, \eta_0)}
  Q(\theta; \theta^{(t)}) 
  = \argmax_{\theta = (V, \Phi, \eta_0)}
  \sum_{i=1}^n \sum_{l=1}^m \log f_{\theta}(X_i, \Lambda_{il}) w_{il}^{(t)}
  \end{equation}
  The M-step is then easily seen to be a quasi-GLM weighted regression of $X$
  on the cubature nodes $\Lambda_{il}$. In particular, we estimate the
  dispersion parameters using weighted Pearson residuals,
  \begin{equation}\label{eq:phi_pearson}
  \hat{\phi}_j^{(t+1)} = \frac{1}{n m} \sum_{i=1}^n \sum_{l=1}^m
  \frac{(X_{ij} - \hat{\mu}_{ijl}^{(t+1)})^2}{w_{ij} w_{il}^*
  \mathcal{V}(\hat{\mu}_{ijl}^{(t+1)})},\quad j\in [p],
  \end{equation}
  where $w_{il}^*$ are the cubature weights at the last EM iteration.
\end{description}

The algorithm for this EM optimization is summarized in Algorithm~\ref{Algo:EM}.
While this computational scheme is robust and efficient, both its complexity
and approximation errors are proportional to the rank~$q$.
It is common in the literature~\citep{2020frank} to assume that $q \ll p$
so that the factorization is parsimonious. For cases when we need a
larger $q$, we explore two alternative optimization methods utilizing
stochastic gradient descent (SGD) next.
\begin{algorithm}
\Data{$X \in \mathbb{F}^{n\times p}$ for some field $\mathbb{F}$}
\Input{Factorization rank $q$, number of Gauss-Hermite cubature nodes $m$,
and max iteration $T$}

Initialize $(V_0, \eta_0)$ using centered DMF or SVD and then $\phi_j$ using Eq~\eqref{eq:phi_pearson} \\
\For{t=$1,\ldots,T$}{
  \Comment{E-step:}
  \For{i=$1,\ldots,n$}{
    Compute Laplace approximation $\Lambda_i | X_i; \theta^{(t)} \approx
    N(\hat{\Lambda}_i, \hat{H}_i^{-1})$ using Fisher scoring
    in~\eqref{eq:em_elli}\\
    Compute Gauss-Hermite nodes $\Lambda_{il}$ and weights $w_{il}^{(t)}$
    for $l=1,\ldots,m$ as defined in~\eqref{eq:emopti_gq}\\
  }
  \Comment{M-Step:}
  Obtain $V^{(t+1)}$ and $\eta_0^{(t+1)}$ by weighted quasi-GLM regression
  in~\eqref{eq:emopti}\\
  Obtain dispersions $\phi_j^{(t+1)}$ for $j=1,\ldots,p$ via Pearson residuals using~\eqref{eq:phi_pearson}
}
Identify $\tilde{V}_{T}, \tilde{\eta}_{T} $ according to Section~\ref{subsec:identi}\\
\Output{MLE estimators $\hat{V} = \tilde{V}_T, \hat{\eta}_0 = \tilde{\eta}_T, \hat{\Phi} = \Phi_T $}
\caption{\textbf{Approximate EM (recommended for small $q$})}\label{Algo:EM}
\end{algorithm}

\subsection{SGD optimization with large rank}
Under some regularity conditions to allow the exchange of differentiation and
integration, the gradient of the likelihood can be written as:
\begin{equation}
\label{eq:grad}
  \begin{aligned}
   \nabla_{\theta} \sum_{i=1}^n \log f_\theta(X_i)
   &= \sum_{i=1}^n \frac{\nabla_{\theta} \int_{\Lambda_i} f_\theta(X_i, \Lambda_i) d\Lambda_i}{f_{\theta}(X_i)}
   = \sum_{i=1}^n \int_{\Lambda_i} \frac{\nabla_\theta f_\theta(X_i,\Lambda_i)}{f_\theta(X_i)} d \Lambda_i\\
   &=\sum_{i=1}^n  \int_{\Lambda_i} \nabla_\theta [\log f_\theta(X_i,\Lambda_i)] \frac{f_\theta(X_i, \Lambda_i)}{f_\theta(X_i)} d\Lambda_i
   =\sum_{i=1}^n  \Exp_{\Lambda_i|X_i; \theta}\big[\nabla_\theta \log f_\theta(X_i, \Lambda_i) \big]
  \end{aligned}
\end{equation}
If we can evaluate Eq~\eqref{eq:grad} efficiently and accurately, we could then
update the EFM parameters through gradient descent with step size $\alpha_t$:
\begin{equation}
\label{eq:gdpre}
  \theta^{(t+1)} = \theta^{(t)} - \alpha_t \nabla_{\theta} \Bigg[
  -\sum_{i=1}^n \log f_\theta(X_i) \Bigg]
  = \theta^{(t)} - \alpha_t \Bigg[
  -\sum_{i=1}^n \Exp_{\Lambda_i\,|\,X_i; \theta}
  \big[\nabla_{\theta} \log f_\theta(X_i, \Lambda_i)\big] \Bigg].
\end{equation}
The gradients $\nabla_\theta \log f_\theta(X_i, \Lambda_i)$ are available in
closed form given the specification of our model in Eq~\eqref{eq:generate};
we derive them in the Appendix. But the posterior expectation
in~\eqref{eq:gdpre} depends on the parameters $\theta$; in practice we adopt a
generalized EM (GEM) approximation with
\[
\nabla_{\theta} \log f_\theta(X; \theta^{(t)}) =
\sum_{i=1}^n \Exp_{\Lambda_i\,|\,X_i; \theta^{(t)}}
\big[\nabla_\theta \log f_\theta(X_i, \Lambda_i) \big].
\]
This approach is equivalent to a gradient descent step on
$-Q(\theta; \theta^{(t)})$ in~\eqref{eq:emopti_estep} since
\[
\nabla_\theta Q(\theta; \theta^{(t)}) =
\nabla_\theta \sum_{i=1}^n \Exp_{\Lambda_i\,|X_i; \theta^{(t)}}
  \big[\log f_\theta(X_i, \Lambda_i)\big] =
  \nabla_{\theta} \log f_\theta(X; \theta^{(t)})  
\]
by Leibniz's rule. A similar, but more computationally expensive, approach is
to perform a Newton-Raphson update~\citep{lange1995gradient,lange1995quasi}.
The EM gradient descent update is then
\begin{equation}
\label{eq:gdem}
  \theta^{(t+1)} = \theta^{(t)} - \alpha_t \nabla_{\theta} \Bigg[
  -\sum_{i=1}^n \log f_\theta(X_i; \theta^{(t)}) \Bigg]
  = \theta^{(t)} - \alpha_t \Bigg[
  -\sum_{i=1}^n \Exp_{\Lambda_i\,|\,X_i; \theta^{(t)}}
  \big[\nabla_{\theta} \log f_\theta(X_i, \Lambda_i)\big] \Bigg].
\end{equation}
In what follows we drop the dependency on $\theta^{(t)}$ to simplify the
notation.

Computing the expectation of the gradient of every observation $i \in [n]$ in
Eq~\eqref{eq:gdem} is, however, expensive. Here we leverage stochastic
optimization to randomly sample partial data to approximate the gradient.
This strategy is termed stochastic gradient
descent~\citep[SGD;][]{robbins1951stochastic} due to its interpretation as a
stochastic approximation of the actual gradient function.
Since the method effectively reduces the sample size by applying a
sub-sampling on the original dataset, the SGD can be used to accelerate all
optimization algorithms with explicit gradient formulation.

Taking the last equality from Eq~\eqref{eq:gdem} and applying the law of large
numbers, we can compute the gradient using stochastic samples of data batch
size $B$ and parameter size $S$: with $X^{(b)}$ sampled from the same law as
the marginal $f_\theta(X_i)$ and $\Lambda^{(s)}$ sampled from the same law as
the posterior $f_\theta(\Lambda_i\,|\,X_i)$,
\begin{equation}
\label{eq:SGDgrad}
\begin{aligned}
  \frac{1}{n} \sum_{i=1}^n \Exp_{\Lambda_i|X_i}\big[\nabla_\theta
  \log f_\theta(X_i, \Lambda_i) \big] & \approx
  \Exp_{X_i} \Bigg\{\Exp_{\Lambda_i|X_i}\big[\nabla_\theta
  \log f_\theta(X_i, \Lambda_i) \big] \Bigg\} \\
  & \approx
  \frac{1}{B} \sum_{b = 1}^B \Exp_{\Lambda_i\,|\,X_i = X^{(b)}}
  \big[\nabla_\theta \log f_\theta(X^{(b)}, \Lambda_i) \big].
\end{aligned}
\end{equation}
In practice, $X^{(b)}$ is sampled uniformly and with replacement from $X$,
while sampling $\Lambda_i$ requires other approximation strategies that
are discussed shortly below.
The batch size $B$ is chosen to be smaller than sample size $n$, which scales
the original complexity with a factor of $B/n$ per iteration.

As it is compared to second-order optimization such as Newton's method, step
size selection is of crucial importance for stochastic gradient descent. A
large step size will make the algorithm oscillate while a small step size
hardly improves our likelihood function. The theoretical analysis states that
we should choose the step size according to the conditional number of the
Hessian matrix~\citep{1999convex}. When the Hessian matrix is not
available, recent researchers have refined the step size selection by
utilizing the momentum and the scale of the parameters. Specifically,
AdaGrad~\citep{2011adagrad} scales the gradient update with the gradient's
second moment while RMSProp~\citep{2012rmsprop} employed exponential decay to
smooth out the gradient direction. More recently, combining both RMSprop and
AdaGrad, the Adam method~\citep{2015adam} has gained well-deserved
attention in modern stochastic optimization research (see, e.g., the review
by~\citet{bottou2018optimization}, including complexity analysis).

Here we adopt the Adam method with the recommended parameters in the original
reference: $\beta_1 =0.9$, $\beta_2 = 0.999$, and $\epsilon = 10^{-8}$.
To avoid the oscillation around the minimum we set a baseline learning rate
$\alpha_0 := \alpha$ and employ a decay learning rate with
$\alpha_t = \alpha/(1 + 0.5t)$ at the $t$-th iteration. When
necessary, the tuning on the hyperparameter $\alpha$ can be conducted by
randomly sampling $\alpha$ on a log grid.
Recently, \citet{zhang2022adam} have shown that Adam is guaranteed to converge
to a critical point under ``strong growth conditions'', for instance when the
learning rate is adjusted adequately. Our learning rate $\alpha_t$ shows good
performance in practice.

To compute the expected gradient in~\eqref{eq:SGDgrad} we proceed as in
Section~\ref{ssec:em} and approximate
$\Lambda_i\,|\,X_i = X^{(b)} \approx N(\hat{\Lambda}_b, \hat{H}_b^{-1})$ with
$\hat{\Lambda}_b$ the posterior mode and $\hat{H}_b$ the negative
expected Hessian at $X^{(b)}$. However, adopting a Gauss-Hermite cubature
to approximate the expectation numerically,
\[
\Exp_{\Lambda_i\,|\,X_i = X^{(b)}} \big[ \nabla_\theta \log f_\theta(X^{(b)},
\Lambda_i) \big]
\approx \sum_{l=1}^m \nabla_\theta \log f_\theta(X^{(b)}, \Lambda_{bl}) w_{bl},
\]
where $\Lambda_{bl}$ and $w_{bl}$ are the cubature nodes and weights relative
to $N(\hat{\Lambda}_b, \hat{H}_b^{-1})$, is still too expensive
computationally. Next, we turn to two alternative strategies.


\subsubsection{Expected gradient via Laplace approximation}
Our first strategy is to prioritize computational efficiency and just rely on
the posterior mode of $\Lambda_i\,|\,X_i = X^{(b)}$. As shown
by~\citet{1989fully},
\begin{equation}
\label{eq:laplderive}
\Exp_{\Lambda_i\,|\,X_i = X^{(b)}} \big[ \nabla_\theta \log f_\theta(X^{(b)},
\Lambda_i) \big]
= \nabla_\theta \log f_\theta(X^{(b)}, \hat{\Lambda}_b)(1 + O(1/p)).
\end{equation}
Therefore, taking
$\Exp_{\Lambda_i\,|\,X_i = X^{(b)}} \big[ \nabla_\theta \log f_\theta(X^{(b)},
\Lambda_i) \big] \approx \nabla_\theta \log f_\theta(X^{(b)},
\hat{\Lambda}_b)$ incurs in a $O(1/p)$ proportional error that should be small
for large $p$ and as $\theta$ converges to the MLE. Following~\eqref{eq:grad}
and~\eqref{eq:SGDgrad} we then have
\begin{equation}
\label{eq:laplgradapprox}
\frac{1}{n} \nabla_\theta \log f_\theta(X) \approx
\frac{1}{B} \sum_{b=1}^B \nabla_\theta \log f_\theta(X^{(b)}, \hat{\Lambda}_b).
\end{equation}

\subsubsection{Expected gradient via Monte Carlo simulation}
When $p$ is of moderate dimension the error in~\eqref{eq:laplderive} can be
significant, so we need to better capture the posterior
$\Lambda_i\,|\,X_i = X^{(b)}$ when evaluating the expectation.
One approach is to perform Monte Carlo simulation,
\begin{equation}
\label{eq:psderive}
\Exp_{\Lambda_i\,|\,X_i = X^{(b)}} \big[ \nabla_\theta \log f_\theta(X^{(b)},
\Lambda_i) \big] \approx \frac{1}{S} \sum_{s=1}^S
\nabla_\theta \log f_\theta(X^{(b)}, \Lambda_b^{(s)})
\end{equation}
where each $\Lambda_b^{(s)} \sim \Lambda_i\,|\,X_i = X^{(b)}$ for $s = 1,
\ldots, S$. This approach is then equivalent to Monte Carlo
EM~\citep[MCEM;][]{wei1990monte}, where the samples are usually obtained via
Markov chain Monte Carlo. Moreover, as indicated in~\citep{caffo2005ascent},
this MCEM for gradient-based optimization often requires an adaptive change of
the sample size $S$ to converge successfully. \citet{2006emadam} circumvents
the need for an adaptive $S$ by averaging over past iterations. The average is
oftentimes weighted with an emphasis on the recent iterations, which is in
fact equivalent to the step size selection of Adam
optimization~\citep{2015adam}.

A good compromise, however, in terms of simplicity and thus efficiency, is to
sample from the approximate posterior: with
$\Lambda_b^{(s)} \sim N(\hat{\Lambda}_b, \hat{H}_b^{-1})$ in~\eqref{eq:psderive}
we have
\begin{equation}
\label{eq:psgradapprox}
\frac{1}{n} \nabla_\theta \log f_\theta(X) \approx
\frac{1}{S \cdot B} \sum_{b=1}^B \sum_{s=1}^S
\nabla_\theta \log f_\theta(X^{(b)}, \Lambda^{(s)}_b).
\end{equation}
Since we are implicitly assuming that the shape of the posterior is determined
solely by a standard Gaussian prior and the likelihood only contributes to the
location and scale, this approximation should be reasonably robust, as claimed
in~\citep{2009inla}. For later reference we denote this approach as
\emph{posterior sampling}.

Finally, we note that MCEM has been shown to have better convergence
properties when compared to SML~\citep{2003MCEMcomp}. Moreover, because this
sampled gradient is proposed to maximize the true likelihood instead of the
simulated likelihood, the sample size $S$ required to compute the expected
gradients does not need to grow in the order of the actual sample size,
e.g. $S = O(\sqrt{n})$ as required for SML~\citep{1995smlsize}.

We summarize these two SGD approaches in Algorithm~\ref{Algo:SGD}

\begin{algorithm}[H]
\Data{$X \in \mathbb{F}^{n\times p}$ for some field $\mathbb{F}$}
\Input{Batch size $B$, sample size $S$, learning rate $\alpha$, rank $q$, and
max iteration $T$; default Adam parameters as $\beta_1 = 0.9$,
$\beta_2 =0.999$, $\epsilon =10^{-8}$.}

\emph{Initialize} $V_0, \eta_0$ using centered DMF or SVD, $\Phi_0$ through
Pearson residuals\\
\emph{Set} $m_0 = \mathbf{0}$ (first moment) and $v_0 = \mathbf{0}$
(second moment)\\

\For{$t=1,\ldots,T$}{
  Sample with replacement batch $X^{(B)}$ from data $X$\\
  \For{$b=1,\ldots,B$}{
    Compute $\hat{\Lambda}_b$ and $\hat{H}_b^{-1}$ as the posterior mode and
    precision respectively of $\Lambda_i\,|\,X_i = X^{(b)}; \theta^{(t)}$ as
    defined in Eq~\eqref{eq:em_elli}\\
    \If(\Comment*[f]{posterior sampling?}){$p$ is \textbf{not} large}{
      \For{$s=1,\ldots,S$}{
        Sample $\Lambda_b^{(s)} \sim N(\hat{\Lambda}_b, \hat{H}_b^{-1})$
      }
    }
  }
  \If(\Comment*[f]{Laplace approximation?}){$p$ is large}{
    Compute the gradient
    $g_t := \nabla_\theta \log f_{\theta}(X; \theta^{(t)})$ as
    in~\eqref{eq:laplgradapprox}\\
  }
  \Else(\Comment*[f]{posterior sampling}){
    Compute the gradient
    $g_t := \nabla_\theta \log f_{\theta}(X; \theta^{(t)})$ as
    in~\eqref{eq:psgradapprox}\\
  }
  Update $m_t = \beta_1 m_{t-1} + (1 - \beta_1) g_t$\\
  Update $v_t = \beta_2 v_{t-1} + (1- \beta_2) (g_t \circ g_t)$ \\
  Update $m_t = m_t/(1 - \beta_1^t)$ and $v_t = v_t/(1 - \beta_2^t)$ \Comment*[f]{bias correction}\\
  Update
  \[
    \theta^{(t+1)} = \theta^{(t)} - \frac{\alpha}{1 + 0.5t}
    \frac{m_t}{\sqrt{v_t} + \epsilon}
  \]
}
Compute the SVD $V'_{t+1} = UDS^\top$ to identify $\tilde{V}_{T}, \tilde{\eta}_{T} $ according to Section~\ref{subsec:identi}\\
\Output{MLE estimators $\hat{V} = V_T, \hat{\eta}_0 = \eta_T, \hat{\Phi} = \Phi_T$}
\caption{\textbf{Adam SGD (recommended for large $q$)}}\label{Algo:SGD}
\end{algorithm}


\subsection{Recovering factors}
\label{ssec:factors}
While factor analysis is traditionally focused on the loadings $V$, we are
also interested in estimating the factors $\Lambda_i$, $i \in [n]$. To this
end, we now turn to estimate $\Lambda_i$ from its posterior conditional based
on the model summarized in~\eqref{eq:generate} with $\theta = (\hat{V},
\hat{\eta}_0, \hat{\Phi}) = \hat{\theta}$ as obtained from~\eqref{eq:opti},
that is,
\[
\hat{\Lambda} = \argmax_{\Lambda_i \in \mathbb{R}^q, i \in [n]}
\sum_{i=1}^n \log f_{\Lambda_i}(\Lambda_i \,|\, X_i, \hat{V}, \hat{\eta}_0, \hat{\Psi}) 
= \argmax_{\Lambda_i \in \mathbb{R}^q, i \in [n]}
\sum_{i=1}^n \log f_{\hat{\theta}}(X_i \,|\, \Lambda_i) + \log f(\Lambda_i),
\]
where $f_{\hat{\theta}}$ is defined as in~\eqref{eq:quasi_family} and $f$ is
the density of a standard multivariate normal. Thus, each $i$-th factor can be
estimated independently and in parallel via a regularized (penalized)
generalized linear (quasi-likelihood) model that regresses $X_i$ on $\hat{V}$
with offsets $\hat{\eta}_0$ and dispersion $\hat{\Phi}$:
\begin{equation}
\hat{\Lambda}_i = \argmax_{\Lambda_i \in \mathbb{R}^q}
\log f_{\hat{V}, \hat{\eta}_0, \hat{\Phi}}(X_i \,|\, \Lambda_i) - \frac{1}{2} \Lambda_i^\top \Lambda_i, \quad i \in [n].
\label{eq:recoverfactor}
\end{equation}

\section{Simulation and Case Studies}\label{sec:example}
In this section, we first present the results of simulation experiments,
demonstrating that our SGD estimation is more effective and superior to SML
estimation. We then benchmark datasets from computer vision and network
analysis to compare our EFM factorization results with other
commonly used methods including non-negative matrix
factorization~\citep[\textbf{NMF},][]{1999nmf}, t-distributed stochastic
neighbor embedding~\citep[\textbf{t-SNE},][]{hinton2002stochastic}, and
deviance matrix factorization~\citep[\textbf{DMF},][]{2021dmf}.
The factorization ranks $q$ for these empirical datasets are determined
based on the rank determination approach proposed in~\citep{2021dmf}.
\subsection{Simulated data}
To assess the effectiveness of our optimization algorithm we applied it to
several simulated datasets. Specifically, we designed small datasets in which
the marginalized likelihood can be computed using SML with a large sample
size, denoted by~$S$. To prevent confusion we use $S$ to refer to the sample
size for gradient evaluation and $R$ to refer to the sample size for marginal
likelihood evaluation. We clarify this notation further in the marginal
likelihood evaluation below:
\begin{equation}
\label{eq:smlloss}
\mathcal{L}(V) = \sum_{i=1}^n \log f(X_i| V)
\approx \sum_{i=1}^n \log \sum_{r=1}^R \frac{1}{R} f(X_i|\Lambda_i^{(r)}, V)
= \sum_{i=1}^n \bigoplus_{r=1}^R \log f(X_i| \Lambda_i^{(r)}, V) - n\log R,
\end{equation}
where $\oplus(x, y) = \log(e^x + e^y)$ is the sum operator in log space.
Note that the evaluation of Eq~\eqref{eq:smlloss} is not required for the
implementation of our Algorithm~\ref{Algo:SGD}. The evaluation is conducted
solely to compare the effectiveness and efficiency of different optimization
methods in minimizing the integrated negative log-likelihood. For non-Gaussian
data likelihoods the integrated negative log-likelihood can only be
accurately computed using SML with large $R$.

To compare the quality of our optimization algorithm with SML we
conducted experiments using simulated data from negative binomial
$(\phi = 20)$, binomial, and Poisson distributions, each with their canonical
link functions. We set the parameters as follows: $n = 500$, $B = 128$,
$q = 2$, $p = 10$, and $\alpha = 0.5$. The sample size $n$ was intentionally
kept small to enable efficient evaluation of the loss function via Monte Carlo
simulation.

Using the true generating parameter $V^*$ we found empirically that
accurate likelihood evaluation using~\eqref{eq:smlloss} requires $R = 1,500$,
which is three times the sample size $n = 500$. This aligns with existing
literature but is still notable, as previous discussions usually focus on
the asymptotic efficiency with a lower bound of $R > \sqrt{n}$
\citep{1995smlsize}. In practice, the required number of Monte Carlo samples
for accurate gradient or likelihood evaluation depends on the
variance of the gradient and likelihood of the dataset, and there is no strict
upper bound. For our experiments we adopt $R = 1,500$ to reliably track the
decrease in loss per iteration from a common starting point. For a fair
comparison, the time required for likelihood evaluation with sample size $R$
is subtracted from the overall optimization time.

To examine how dimensionality and variance affect different optimization
algorithms, we first performed two experiments with $p = 5$ and $p = 10$,
followed by two additional experiments with a larger dimension size of
$p = 512$. The smaller values of $p$ were chosen to ensure numerical stability
for SML optimization, which, as discussed in
Section~\ref{subsubsec:sml_problem}, encounters evaluation issues when $p$ is
large. For each experiment, we fix both the initialization and random seed to
enable fair comparison of the optimization paths, and systematically varied
the sample size $S$ in~\eqref{eq:sml_grad}.

\begin{figure}[H]
\centering
\includegraphics[width=\textwidth]{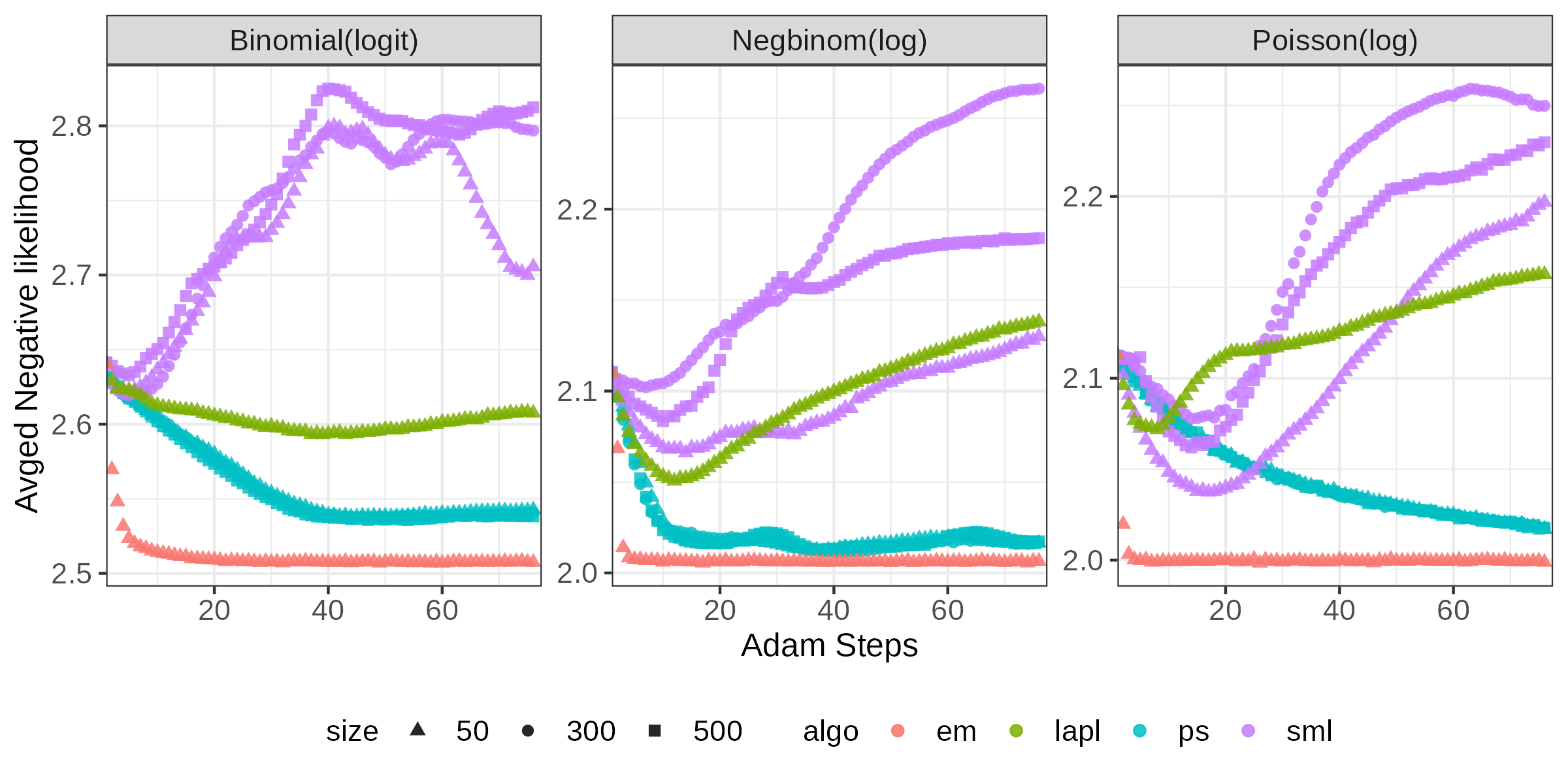}
\caption{EFM optimization path comparison, $p = 5$. EM is
Expectation-Maximization, LAPL is the SGD based on Laplace approximation, PS
is the SGD based on posterior sampling, and SML is simulated maximum
likelihood optimization. The PS and SML sample sizes $S$ are 50, 300, and 500.}
\label{fig:EFM_vs_SML_p5}
\end{figure}

As shown in Figure~\ref{fig:EFM_vs_SML_p5}, the posterior sampling (PS)
optimization method is largely insensitive to the choice of sample size $S$.
In contrast, the optimization paths for SML change considerably as $S$ varies,
with larger values of $S$ yielding faster reductions in the loss function.
The Laplace approximation decreases the loss function at the slowest rate,
primarily due to its gradient approximation error of $O(1/p)$
(see~\eqref{eq:laplderive}). These results suggest that, for problems with
small data dimension $p$, the EM and PS optimization methods are preferred
thanks to their efficiency, reduced sensitivity to sample size $S$, and
superior numerical stability.

In theory, the accuracy of the LAPL optimization improves as the relative
error decreases at the rate $O(1/p)$. To evaluate the performance of LAPL at
moderate data dimensions, we repeated the experiments using $p =10$.
For both SML and PS optimization methods, we set $S = 500$ to directly compare
their efficiency. As before, we tracked the optimization paths by recording
the loss at each Adam optimization step.

\begin{figure}[H]
\centering
\includegraphics[width=\textwidth]{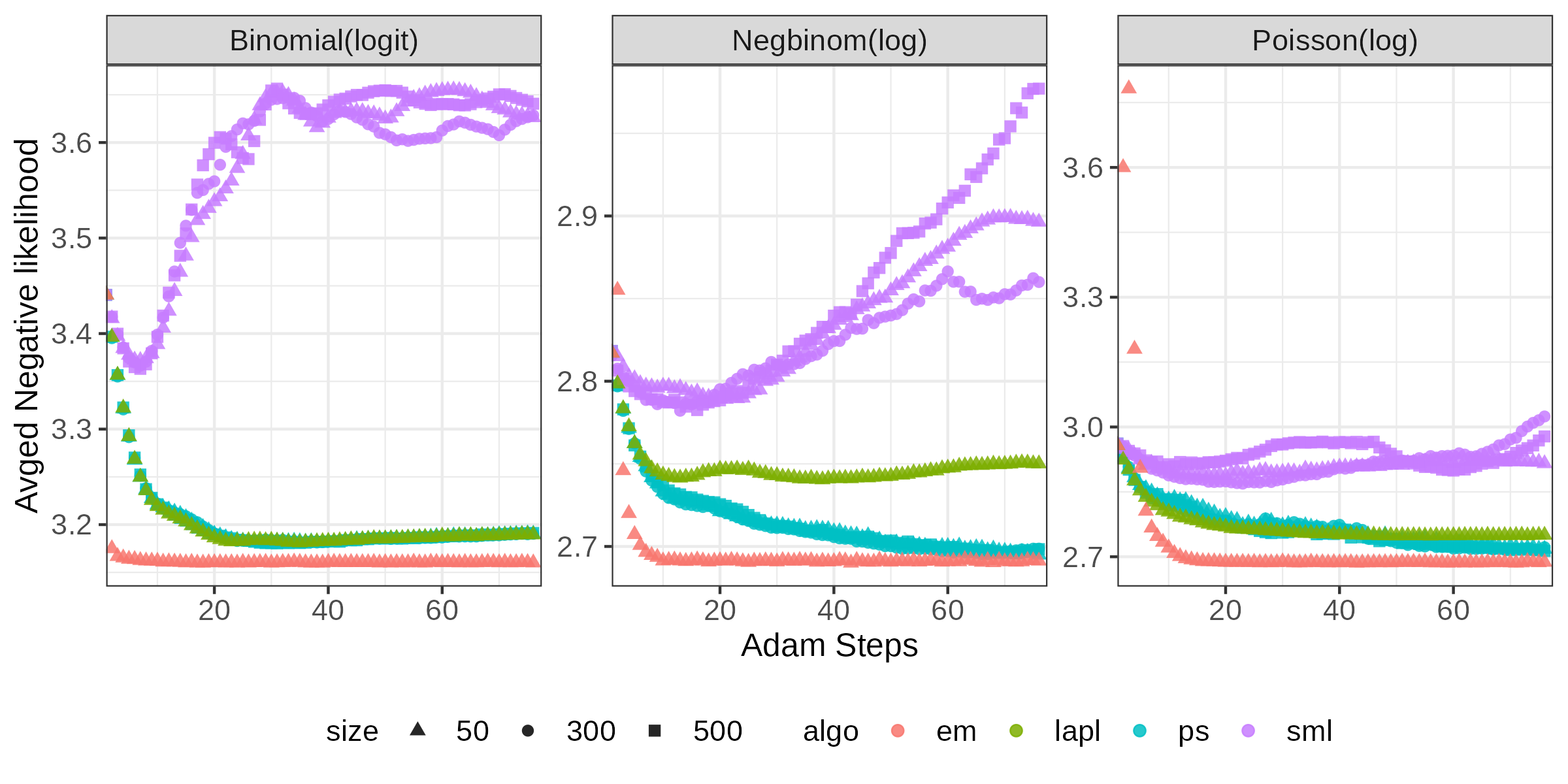}
\caption{EFM optimization path comparison, $p=10$. Algorithms and SML sample
sizes as in Figure~\ref{fig:EFM_vs_SML_p5}.}
\label{fig:EFM_vs_SML_p10}
\end{figure}

As we can see in Figure~\ref{fig:EFM_vs_SML_p10}, posterior sampling
optimization remains the most efficient of the three methods, exhibiting a
steeper decrease in loss per time unit. However, as PS approaches the
convergence region, its optimization shows higher variance compared to LAPL
optimization. This suggests that Laplace optimization may be advantageous when
the gradient evaluation incurs high variance. In contrast, SML
optimization consistently reduces the negative likelihood at the slowest rate,
making it unsuitable for most practical applications.

To further assess optimization performance on high-dimensional data, we
conducted experiments with $p=512$. Under this scenario, SML completely
failed due to numerical instability, consistently either (i) increasing the
loss at each step without convergence or (ii) converging to a higher loss
than EM, LAPL, and PS methods. Consequently, our comparison at this data
dimension focuses only on PS at varying sample sizes $S$ and
LAPL, evaluating their optimization paths as a function of time.

\begin{figure}[H]
\centering
\includegraphics[width=\textwidth]{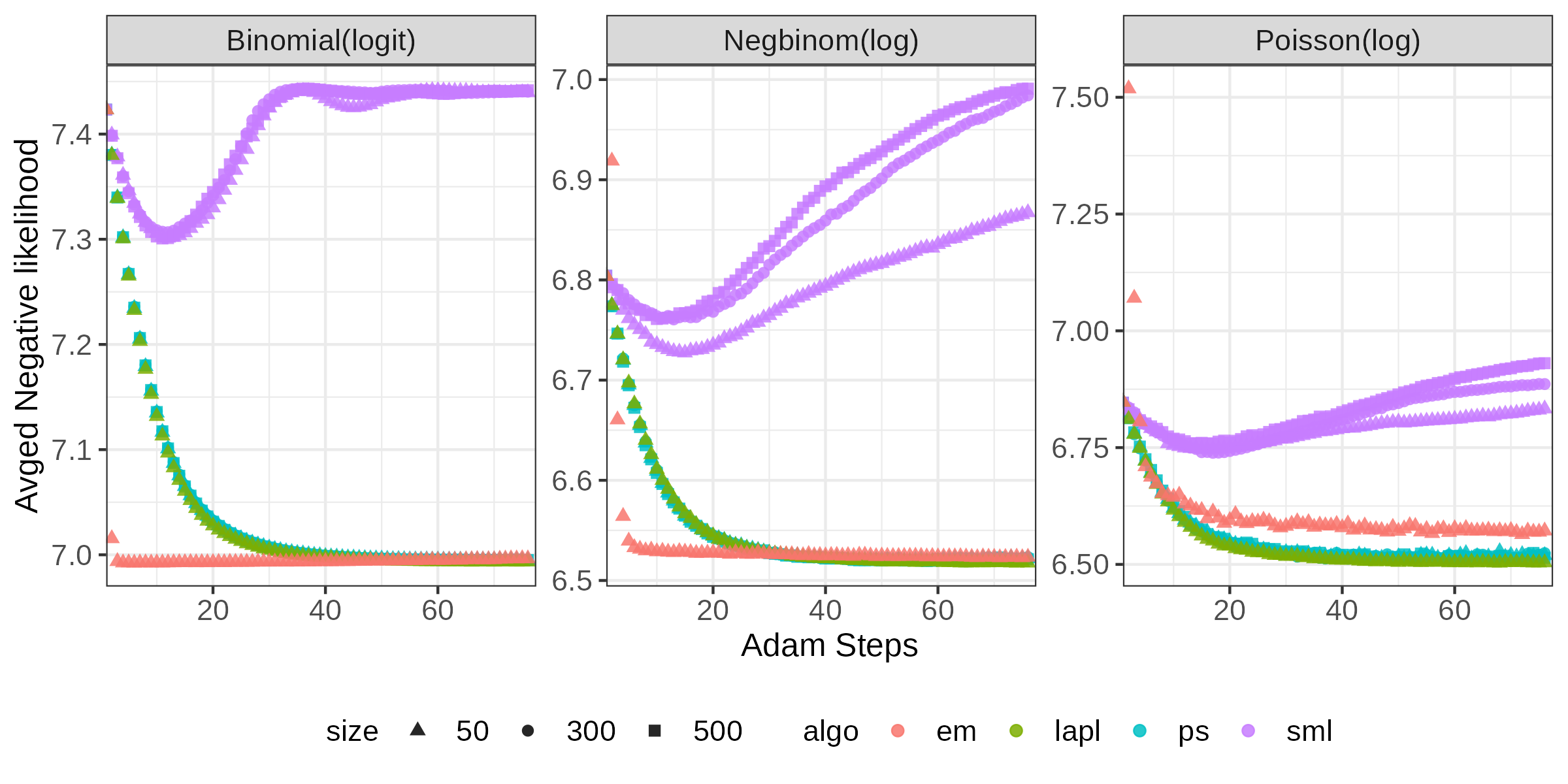}
\caption{EFM optimization path comparison, $p=512$. Algorithms and SML sample
sizes as in Figure~\ref{fig:EFM_vs_SML_p5}.}
\label{fig:EFM_vs_SML_p512}
\end{figure}

As shown in Figure~\ref{fig:EFM_vs_SML_p512}, despite the potentially
larger Monte Carlo error in gradient evaluation at a small sample size
($S = 50$), PS optimization still achieves convergence when compared to LAPL
and EM optimizations. This behaviour is highly desirable, as it demonstrates
that efficient optimization in high dimensions is possible even with
low computational resources. However, as the variance in gradient evaluation
increases, PS optimization may require a larger sample size $S$.
To test this hypothesis, we increased the magnitude of $V^* = U^* D^*$ in
our simulation by multiplying its diagonal elements by a constant $c > 1$,
which amplifies the gradient magnitude according to the relationships
listed in Table~\ref{tab:lossefm}.

\begin{table}[H]
\captionof{table}{Distributions and their respective gradient and Hessian functions.}
{\scriptsize
\begin{center}
\begin{tabular}{lccc} \toprule
Distribution $F$                  & Negative log-Likelihood $l(X)$                                              & Gradient $\nabla_{V_j}l(X)$                                                   & Hessian $\nabla^{2}_{\Lambda_i} l(X)$ \\ \midrule
Gaussian (identity)               & $\frac{1}{2\sigma^2}\sum_{j=1}^p (X_j-\Lambda V_j)^\top(X_j-\Lambda V_j)$   & $-\frac{1}{\sigma^2}\Lambda^\top(X_j-\Lambda V_j^\top)$                       & $\frac{1}{\sigma^2}V^\top V$ \\
Poisson (log)                     & $\sum_{j=1}^p 1_n^\top \exp(\Lambda V_j^\top) - X_j^\top(\Lambda V_j^\top)$ & $-X_j^\top\Lambda + \exp(\Lambda V_j^\top)^\top\Lambda$                       & $V^\top \text{Diag}[\exp(\Lambda_i V^\top)]V$ \\
Gamma (log)                       & $-\phi [\sum_{j=1}^p X_j^\top \Lambda V_j^\top+ \log(-\Lambda V_j^\top)]$   & $(-1 / \Lambda V_j^\top)\Lambda - X_j^\top \Lambda$                           & $V^\top \text{Diag}^2[\phi/\Lambda_i V^\top] V$ \\
Binomial (logit)                  & $\sum_{j=1}^p - (w_j\circ X_j^\top) \Lambda V_j^\top +$                     & $[w_j/(1+\exp(-\Lambda V_j^\top))]^\top\Lambda-$                              & $V^\top (w_i \circ \text{Diag}[\exp (\Lambda_i V^\top)$ \\
                                  & $w_j^\top \log(1_n+\exp(\Lambda V_j^\top))$                                 & $(w_j \circ X_j)^\top\Lambda$                                                 & $/(1+ \exp(\Lambda_i V^\top a))])V$ \\
Negative binomial ($\alpha$; log) & $\sum_{j=1}^p -X_j^\top [\Lambda V_j^\top -$                                & $-X_j^\top\Lambda + $                                                         & $V^\top \text{Diag}[\exp(\Lambda_i V^\top) +$ \\
                                  & $\alpha 1_n^\top \log(1_q- \exp(\Lambda V_j^\top))]$                        & $\alpha\exp(\Lambda V_j^\top)^\top/(1_n^\top -\exp(\Lambda V_j^\top))\Lambda$ & $\alpha \exp(2\Lambda_i V^\top)]V$ \\
\bottomrule
\end{tabular}
\label{tab:lossefm}
\end{center}
}
\end{table}

\begin{figure}[H]
\centering
\includegraphics[width=\textwidth]{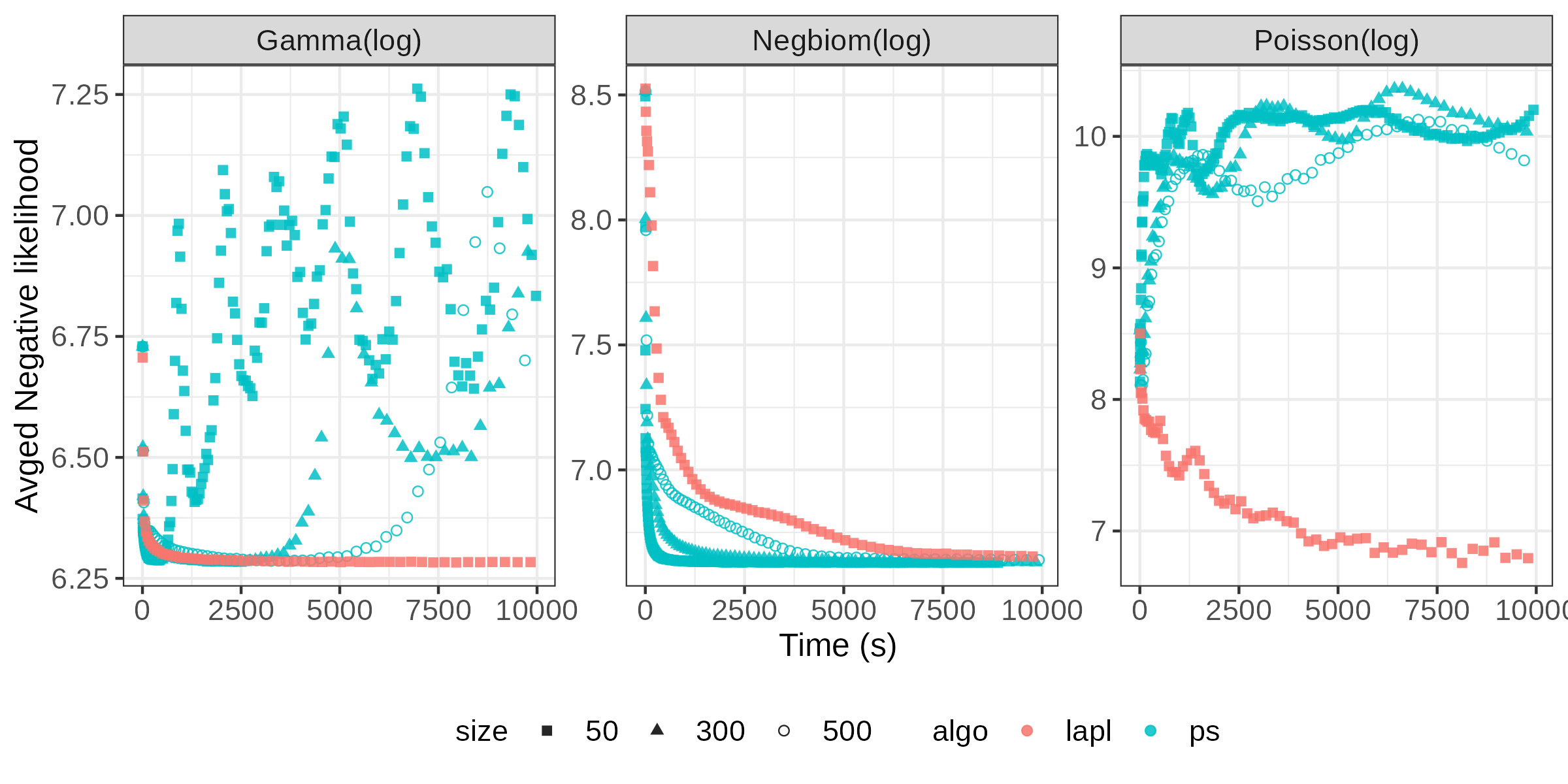}
\caption{EFM optimization path comparison for SGD between Laplace
approximation (LAPL) and posterior sampling (PS) approaches, $p=512$ and
larger $V$ diagonal. The PS sample sizes are 50, 300, and 500.}
\label{fig:EFM_vs_SML_p512LV}
\end{figure}

We continued our data simulation with a large dimension $p = 512$ to
compare PS and LAPL performances. As shown in
Figure~\ref{fig:EFM_vs_SML_p512LV}, when the sampled gradients exhibit high
variance, PS optimization deteriorates, indicating a greater need for a larger
sample size $S$. In such scenarios---especially with non-Gaussian priors,
higher dimensionality, and increased variance in gradient evaluation---LAPL
optimization is preferable. Nonetheless, given the efficiency PS demonstrated
in earlier simulations, we adopted posterior sampling optimization for
subsequent empirical studies, with careful monitoring of loss reductions at
each iteration.

\begin{figure}[H]
\centering
\includegraphics[width=\textwidth]{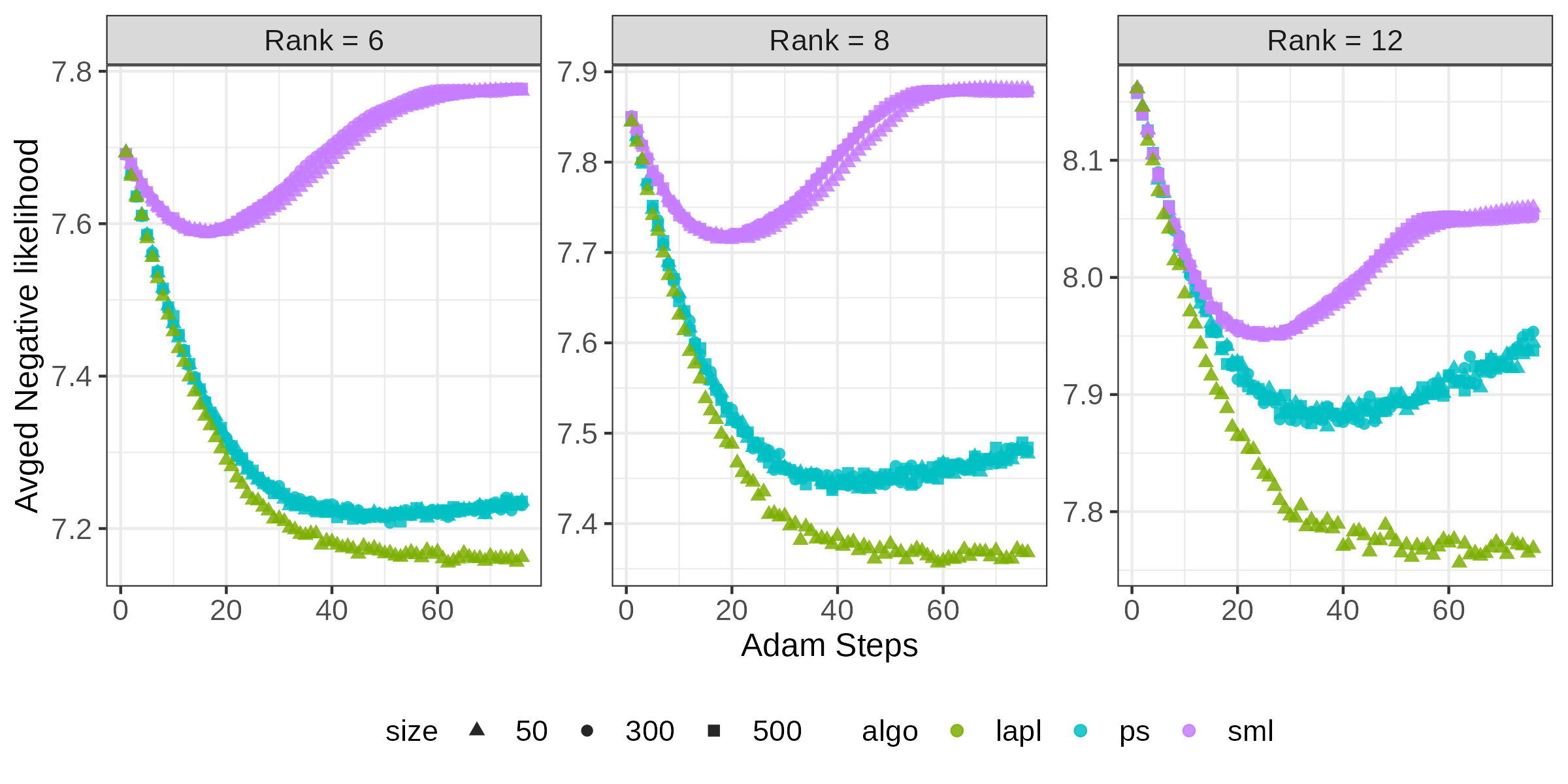}
\caption{EFM optimization path comparison, $p=512$ and large rank $q \in \{6,
8, 12\}$. SGD with Laplace approximation (LAPL), with posterior sampling (PS),
and simulated maximum likelihood (SML). The sample sizes for PS and SML are
50, 300, and 500.}
\label{fig:EFM_vs_SML_p512Lq}
\end{figure}

Although factor modeling is typically performed with small rank $q$,
we conducted additional experiments to demonstrate that our proposed
optimization algorithms outperform SML even at larger ranks.
Using $p= 512$, $n = 500$, and a binomial factorization family, we repeated
the optimization experiments for ranks $q = 6$, $8$, and $12$.
Since our EM algorithm is only recommended for small $q$, we excluded it
and compared only the remaining three algorithms: LAPL, SML, and~PS.
The results of these comparisons are summarized in
Figure~\ref{fig:EFM_vs_SML_p512Lq}.

Finally, to empirically evaluate method performance, we used binomial EFM as
an example and benchmarked computation times per iteration for four different
algorithms across various $p$ and $q$ values. Specifically, for $q = 2$,
we tested $p=5$, $p=10$, and $p=512$; for $p=512$, we evaluated $q=6$, $q=8$,
and $q=12$, omitting the EM method since it is only recommended for small $q$.
The results are listed in Table~\ref{tab:comptime}. Key findings include:
\begin{itemize}
    \item Our SGD PS optimization scales better than SML with respect to both
      large dimension $p$ and large sample size $S$
    \item Among all alternatives, EM requires the longest computational time
      per iteration, but may need fewer total iterations at low rank $q$
      (overall running time not shown).
    \item SGD LAPL optimization delivers the best per-iteration performance
      for large $p$ and $q$.
\end{itemize}

\begin{table}[H]
\caption{Computation time benchmarks (seconds per step) with varying rank $q$
and dimension $p$ for SML, SGD with Laplace approximation (LAPL), SGD with
posterior sampling (PS), and EM. SML and PS running times depend on the sample
size $S$.}
\label{tab:comptime}
\begin{center}
\begin{tabular}{lrrrrrrr}
\toprule
     &     & \multicolumn{3}{c}{$q=2$}  & \multicolumn{3}{c}{$p=512$} \\
\cmidrule(lr){3-5}\cmidrule(lr){6-8}
Method & $S$ & $p=5$  & $p=10$ & $p=512$  & $q=6$ & $q=8$ & $q=12$ \\
\midrule
SML  & 50  & 0.07  & 0.11  & 12.2 & 11.4  & 12.3  &  18.1  \\
     & 300 & 0.26  & 0.39  & 44.0 & 43.7  & 47.0  &  67.8  \\
     & 500 & 0.42  & 0.64  & 67.6 & 78.9  & 72.9  & 110.0  \\
PS   & 50  & 0.24  & 0.46  &  5.6 &  2.1  &  2.8  &   2.7  \\
     & 300 & 0.28  & 0.59  & 19.7 &  7.3  &  8.7  &  10.1  \\
     & 500 & 0.31  & 0.70  & 30.9 & 13.9  & 14.1  &  15.2  \\
LAPL & --- & 0.12  & 0.13  &  1.8 &  1.8  &  1.9  &   4.3  \\
EM   & --- & 1.31  & 1.39  & 16.9 & ---   & ---   & ---    \\
\bottomrule
\end{tabular}
\end{center}
\end{table}

\subsection{Covariance modeling}
\label{ssec:covariance}
The Gaussian factor model is widely used due to its efficiency in covariance
estimation for high dimensional data~\citep{fan2008high}. Using a similar
setup to~\citep{fan2008high}, we simulate three quasi-factor datasets with
$n = 756$, $p \in \{66, 116, \ldots, 466\}$ and four families (quasi-Poisson,
negative binomial, binomial, and Poisson). For comparison, we used the same
prior configuration for both factors $\Lambda$ and projection matrix $V$
according to the setup in Table~1 of~\citep{fan2008high}. That is, for each
$p$ and family:

\begin{enumerate}[(i)]
    \item We first simulate for $i\in[n], j \in[p]$
\begin{equation}
    \begin{aligned}
        \Lambda_i &\sim N\begin{bmatrix}
\begin{pmatrix}
0.023558\\
0.012989\\
0.020714
\end{pmatrix}\!\!,&
\begin{pmatrix}
1.2507 & 0 & 0\\
0 & 0.31564 & 0\\
0 & 0 & 0.19303
\end{pmatrix}
\end{bmatrix}\\
V_j &\sim N\begin{bmatrix}
\begin{pmatrix}
0.78282\\
0.51803\\
0.41003
\end{pmatrix}\!\!,&
\begin{pmatrix}
0.029145 & 0.023873 & 0.010184\\
0.023873 & 0.053951 & -0.006967\\
0.010184 & 0.006967 & 0.086856
\end{pmatrix}
\end{bmatrix}\\
\Phi_j & \sim  \begin{cases}
      \text{Gamma}(\alpha = 4.0713, \beta =0.1623) & \text{ for quasi-Poisson}\\
      1 & \text{for others}
    \end{cases} \\
w_{ij} & \sim  \begin{cases}
      \text{Poisson}(20) & \text{for binomial}\\
      1 & \text{for others}
    \end{cases}
    \end{aligned}
    \label{eq:jianqing_simu}
\end{equation}
\item Conditional on simulated $(\Lambda, V, \Phi)$, we generate
  $X \in \mathbb{F}^{n\times p}$ using the four quasi-families with
  quasi-densities $f$ satisfying~\eqref{eq:generate}.
\item Based upon generated data $X$, quasi-family defined by density $f$ and
  prior of $\Lambda$ in \eqref{eq:jianqing_simu},
  we estimate $\theta = (V, \Phi, \Lambda|X)$ using maximum likelihood, as
  defined in Eq~\eqref{eq:opti}.
\item We compute covariance estimates using the naive sample covariance
  $\frac{X X^\top}{n-1} - \frac{X 1 1^\top X^\top}{n(n-1)}$ and the total
  covariance $\text{Cov}_{\hat{\theta}}(X)$ by plugging the estimated $\theta$
  into Eq~\eqref{eq:total_cov}.
\item We evaluate covariance estimation error using Frobenius norm
  $\|\Sigma - \hat{\Sigma}\|_F$, entropy loss $\text{tr}(\hat{\Sigma}
  \Sigma^{-1}) - \log|\hat{\Sigma} \Sigma^{-1}| - p$, and normalized
  Frobenius loss $\frac{1}{\sqrt{p}}\| \Sigma^{-1/2}(\hat{\Sigma} - \Sigma)
  \Sigma^{-1/2}\|_F$.
\end{enumerate}

\begin{figure}[H]
\centering
\includegraphics[width=\textwidth]{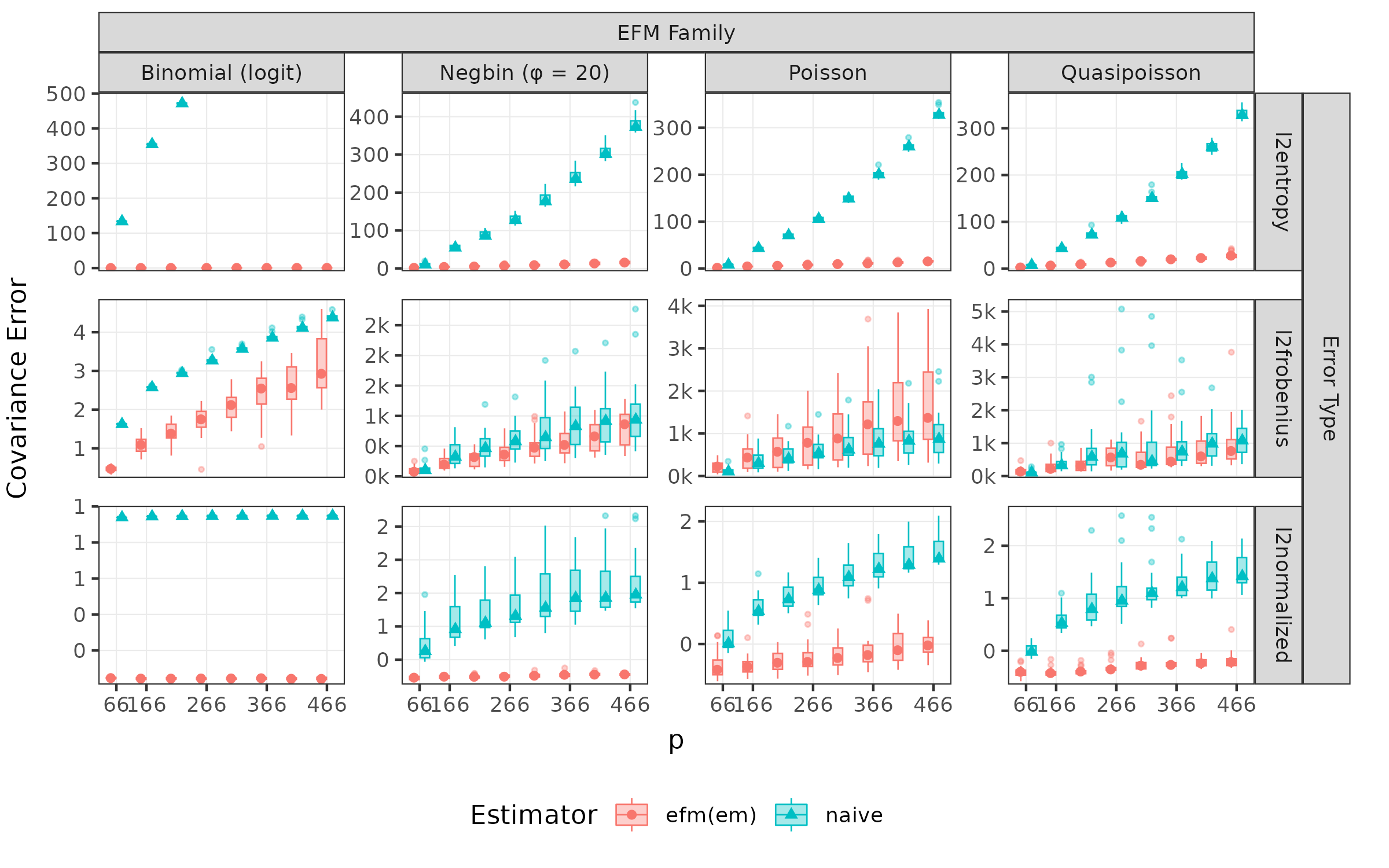}
\caption{Covariance estimation error comparison between EFM EM optimization
and naive covariance estimator. Boxplots are based on $k = 5$ replicates under
the scenarios described in steps (i) through (v).}
\label{fig:cov_error}
\end{figure}

We compared the performance of EFM EM estimates with that of the naive sample
covariance estimator using five replications of steps (i)--(v) above, with
results presented in Figure~\ref{fig:cov_error}. Our findings echo prior
reports for the Gaussian case~\citep{fan2008high}, showing similar errors in
non-Gaussian covariance estimation. Based on normalized Frobenius and entropy
losses, the EM optimization provides much more accurate covariance matrix
estimates for high-dimensional data than the naive estimator.

\subsection{Computer vision data}
To demonstrate the improved representational capabilities of our EFM approach,
we also carried out experiments on computer vision datasets.

\subsubsection{Fashion MNIST dataset}
The MNIST dataset is among the most popular computer vision benchmarks,
containing 70,000 images of handwritten digits labeled 0 to 9.
The Fashion-MNIST dataset~\citep{2017mnist}, which also comprises 70,000
images across ten classes, was introduced to provide a more challenging
classification task that the original MNIST. In this study, we apply our EFM
factorization using a negative binomial family and log link to Fashion-MNIST,
and compare the resulting factors with those from DMF (using the same family
and link) and NMF. As an additional non-parametric baseline, we include
t-distributed stochastic neighbor
embedding~\citep[t-SNE;][]{van2008visualizing}, which is well suited for
visualization and clustering in classification tasks.

\begin{figure}[H]
\centering
\includegraphics[width=.49\textwidth]{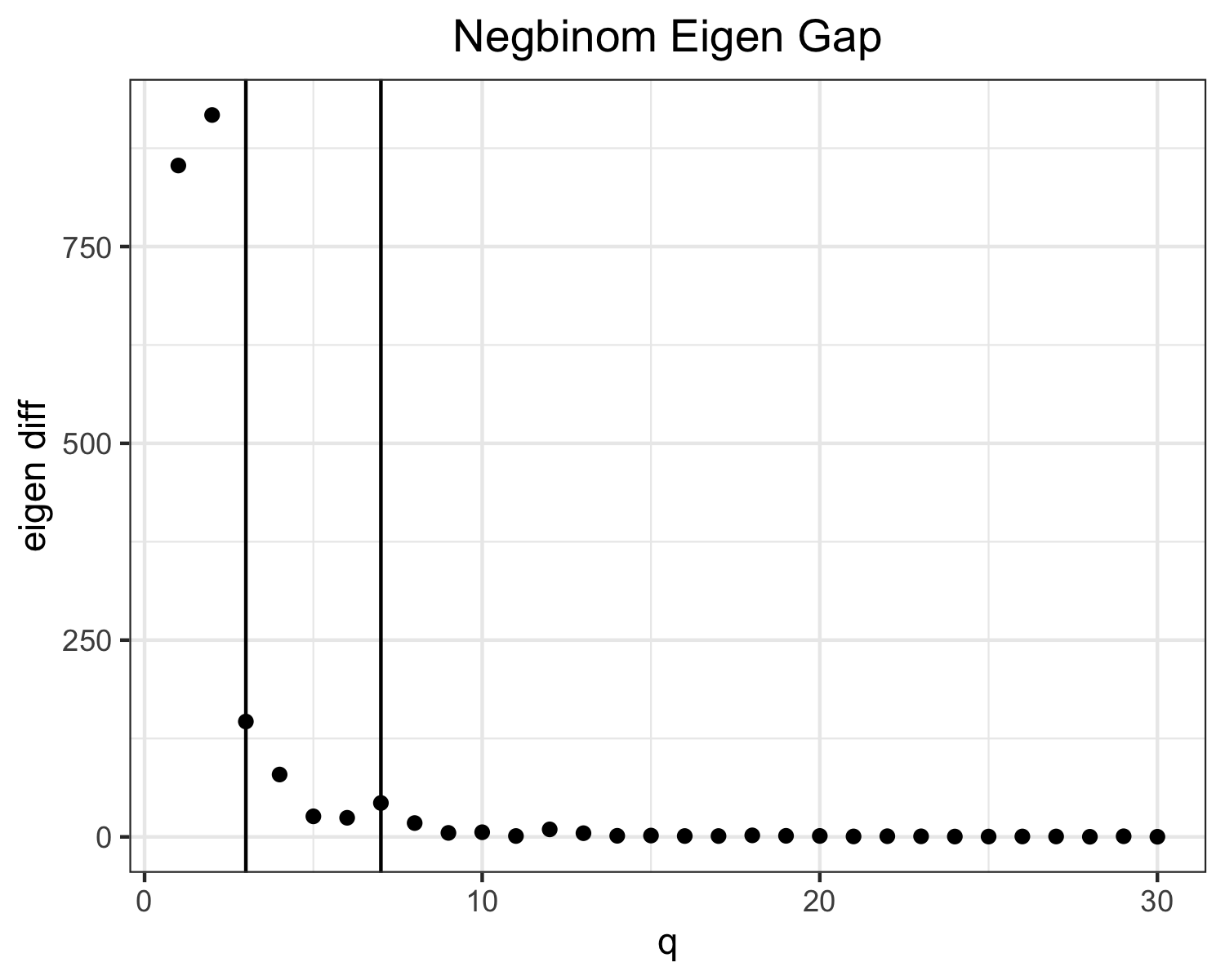}\hfill
\includegraphics[width=.5\textwidth]{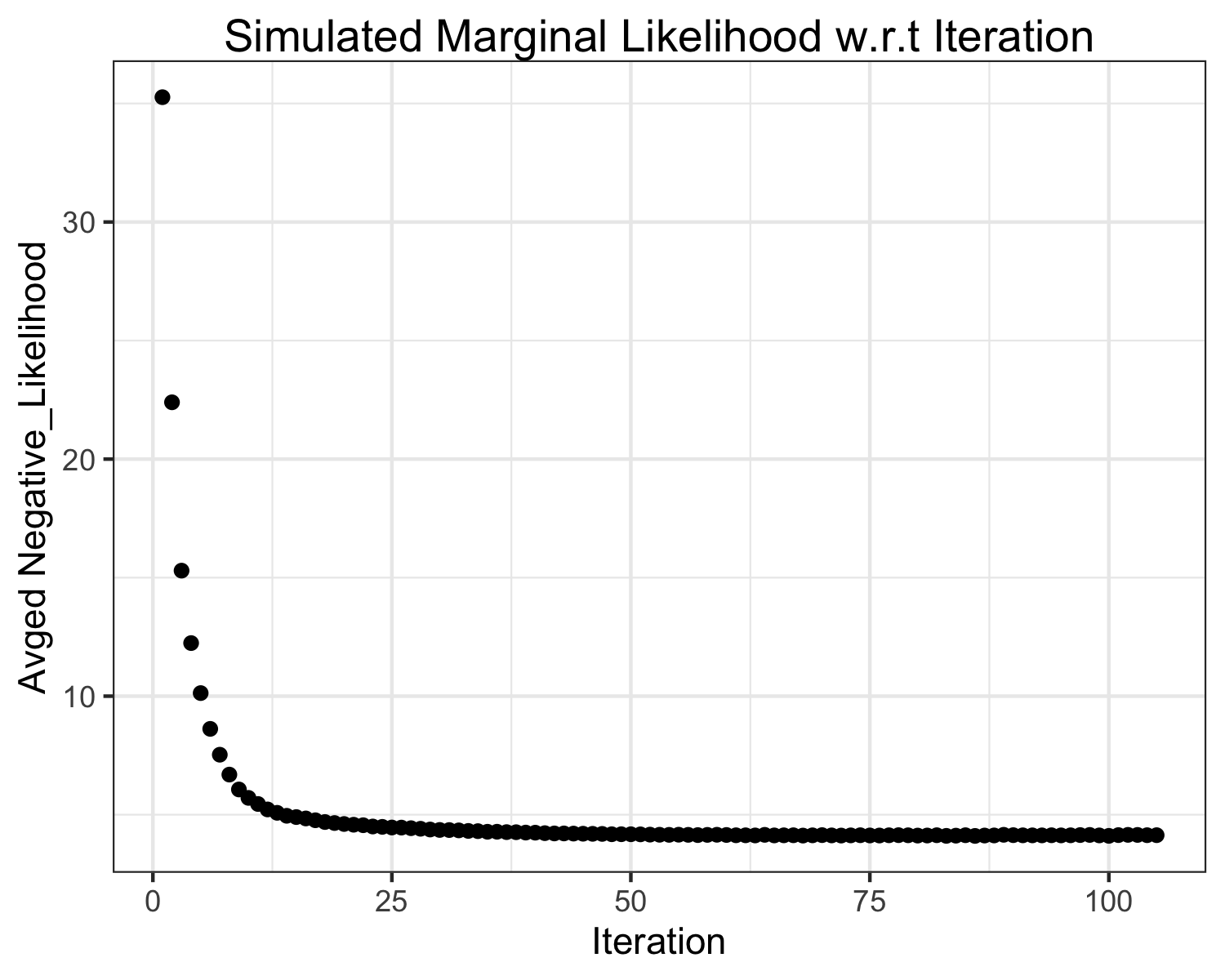}
\caption{Rank determination based on the eigen-gap from a DMF fit for the
Fashion-MNIST dataset (scree plot on the left) and simulated maximum
likelihood convergence assessment (negative likelihood trace on the right).}
\label{fig:mnist_like}
\end{figure}

To determine the rank, we adopted the rank determination procedure suggested
for DMF~\citep{2021dmf}, a more formal version of the scree test.
The resulting scree plot in Figure~\ref{fig:mnist_like} indicates potential
ranks three or seven for the factorization; we adopt rank three for
visualization convenience. The simulated maximum likelihood method achieves
convergence after about 25 epochs with $B =256$, $S = 50$, and $\alpha = 0.5$.

\begin{figure}[H]
    \centering
    \includegraphics[width=.25\textwidth,height =2.3in]{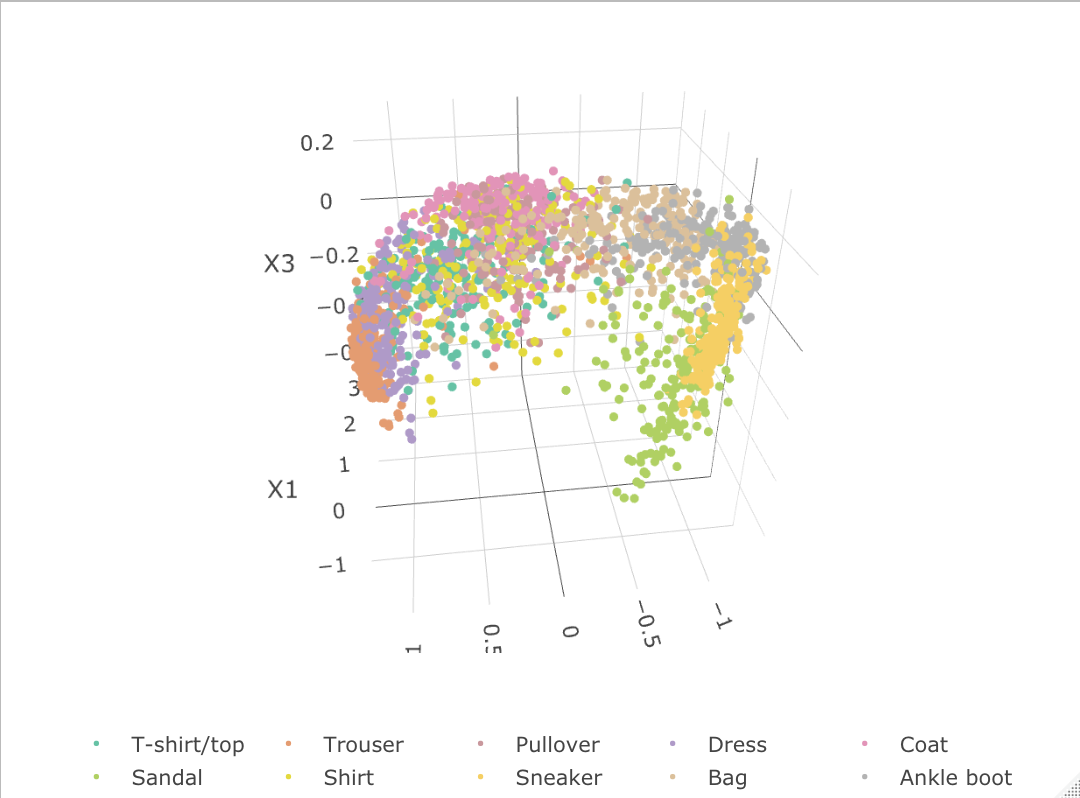}\hfill
    \includegraphics[width=.24\textwidth,height =2.3in]{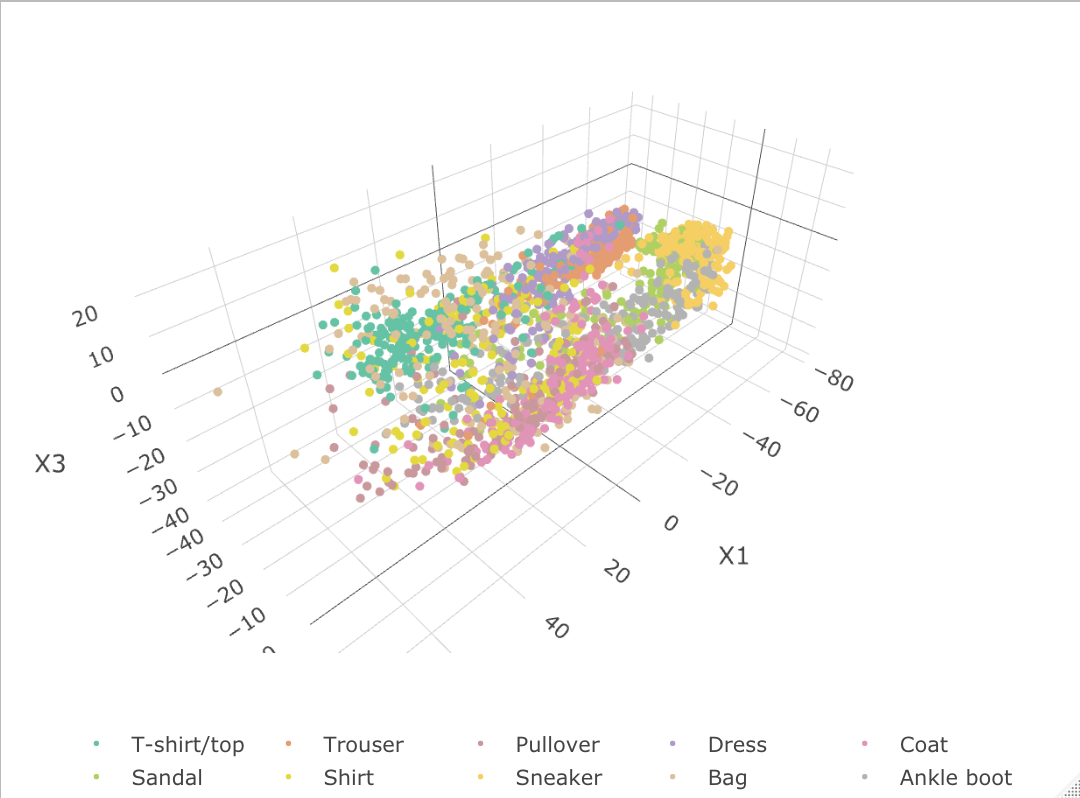}\hfill
    \includegraphics[width=.25\textwidth,height =2.3in]{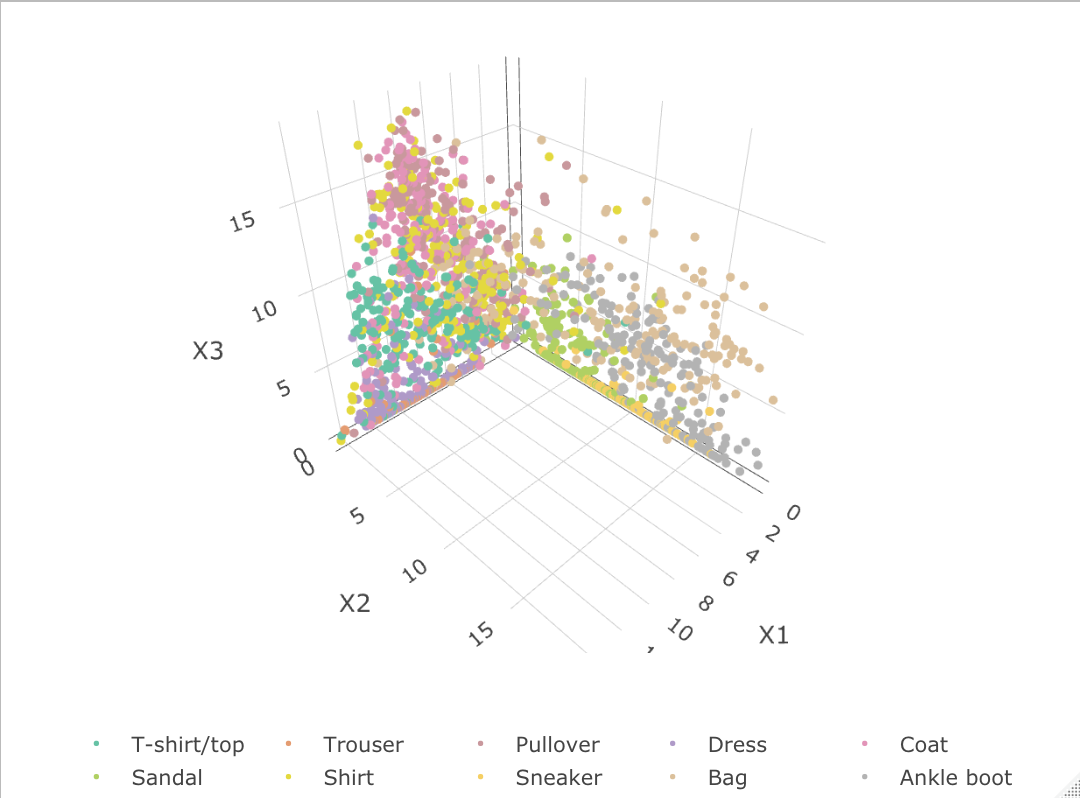}\hfill
    \includegraphics[width=.24\textwidth,height =2.3in]{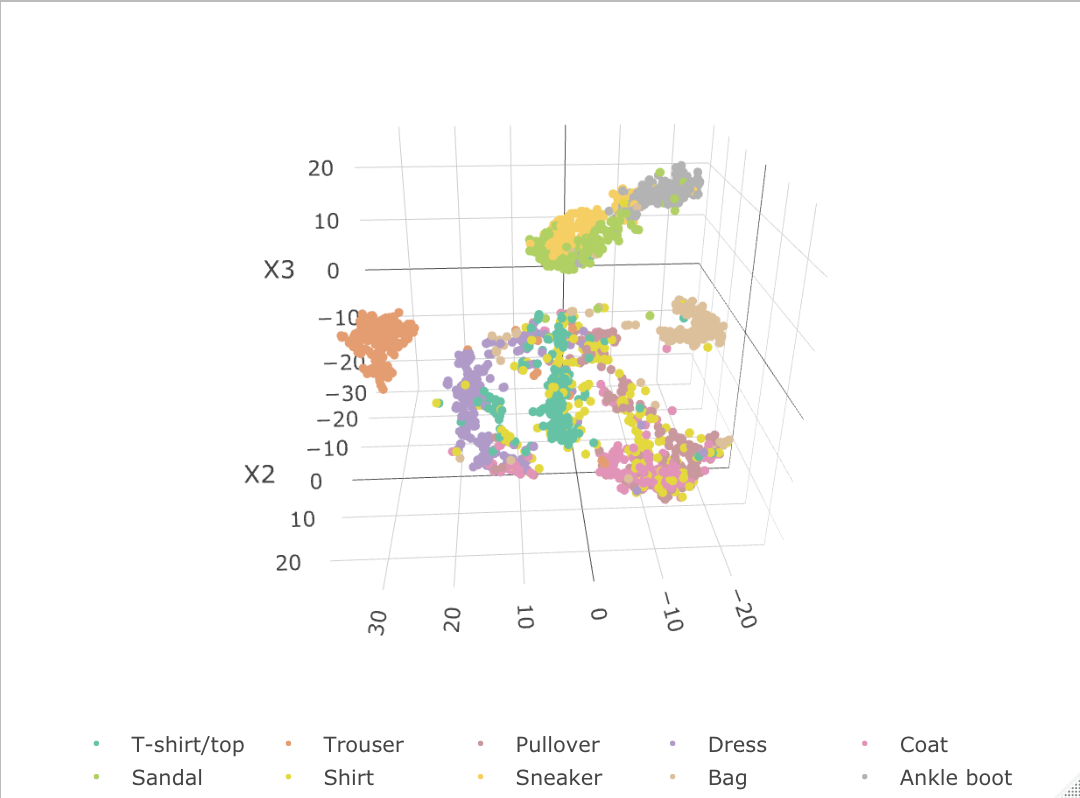}
    \caption{Fashion-MNIST three factors, colored by true classification. Left
    to right panels show results for EFM, DMF, NMF and t-SNE, respectively.}
    \label{fig:cvmnist}
\end{figure}

To illustrate the superior performance under a general model formulation, we
used only the first 2,000 samples of the 70,000 data as our training dataset
to estimate $\hat{V}$. After obtaining this $\hat{V}$, we conduct penalized
GLM regression as described in Section~\ref{ssec:factors} based upon another
2,000 sampled testing set $X$ to estimate $\hat{\Lambda}_i$. These
$\hat{\Lambda}_i$ can thus be considered as an out-of-sample latent estimation
based upon the EFM estimated $\hat{V}$. Due to the factorization algorithm
setup, t-SNE and NMF results are based on the training dataset and the DMF and
EFM results are obtained on the separate testing set. We summarize the
factorization results in Figure~\ref{fig:cvmnist}.

As indicated by the Fashion-MNIST factorization results, both our EFM and the
t-SNE methods show great separability of the 10 fashion classes with some
mistakes on pullover, shirt, and coat, which are actually similar classes
according to their image representations. The NMF performs the worst,
utilizing only two dimensions to separate the 10 different classes. Without
the stochastic optimization and regularization, the DMF performs no better
than our EFM on the testing samples, potentially due to overfitting.

To quantitatively measure the separability of the embeddings, we split the
2,000 factorized embeddings into train/test sets in a 50-50 ratio. Then taking
the factorized embeddings as input, we train and test the classification
performance of three inherently different classification models: decision
trees~\citep{breiman2017classification} using R package~\texttt{rpart},
multinomial regression using R package~\texttt{nnet}, and k-nearest
neighbors~\citep{cover1967nearest} using function~\texttt{knn3} from R package
\texttt{caret}.

Since this is a multiclass classification problem, the performance is
evaluated according to multiclass AUC~\citep{hand2001simple} with results
reported in Table~\ref{tab:mnish10_auc}.
We note that although the t-SNE has the overall best multiclass AUC score, it
is a non-parametric estimation method that is designed for visualization tasks
and thus can better separate classes, serving as a benchmark reference.
Our EFM model has the best AUC score from the factorization methods,
comparable to the t-SNE reference performance.

\begin{table}[H]
\begin{center}
\caption{Comparison of predicted classification of Fashion-MNIST factor embeddings using multiclass AUC.}
\begin{tabular}{cccc}\toprule
\textbf{Embedding/Model} & \textbf{Tree} & \textbf{Multinomial} & \textbf{kNN} \\ \midrule
\textbf{t-SNE}           & 0.92        & 0.95               & 0.95       \\
\textbf{NMF}             & 0.88        & 0.90               & 0.87       \\
\textbf{DMF}             & 0.88        & 0.87               & 0.89       \\
\textbf{EFM}             & 0.90        & 0.94               & 0.92       \\ \bottomrule
\end{tabular}
\label{tab:mnish10_auc}
\end{center}
\end{table}

\subsubsection{ORL face dataset}
Our EFM method also yields reliable estimates with reasonable uncertainty when
used for image restoration. To illustrate this advantage, we also conducted
experiments on the
{ORL face dataset}~\citep{zhu2019multi}, which contains 40 subjects with
pictures taken under 10 different conditions. Each image is in a
$n = 32\times32 = 1024$ dimensional space, a high pixel resolution for
restoration. Each image element is a integer-valued pixel number in the range
$[0,1, \cdots, 255]$ where zero represents the white background color.

We process those $p = 400$ (40 subjects $\times$ 10 conditions) images by
storing each flattened image as a column in our data matrix
$X \in [0,1,\cdots, 255]^{n \times p}$. With such a matrix as input, we
estimate EFM parameters $\hat{\theta} = (\hat{V}, \hat{\eta}_0, \hat{\Phi})$
using Eq~\eqref{eq:opti}.
Here $\hat{V} \in \tilde{\mathcal{S}}_{p,q}(\mathbb{R})$ estimates a (scaled)
latent basis for the 400 human faces. We adopt a negative binomial family for
the EFM with an estimated $\hat{\phi} = 0.651$ by the method of moments. For
comparison, we also fit an Eigenface model~\citep{turk1991face} based on a
singular value decomposition of $X$. We adopt $q = 41$ for both factorizations
since we expect to find 40 individual faces and one ``average face'' and to
favor the SVD with a larger rank.

\begin{figure}[H]
\centering
\includegraphics[width=0.9\linewidth]{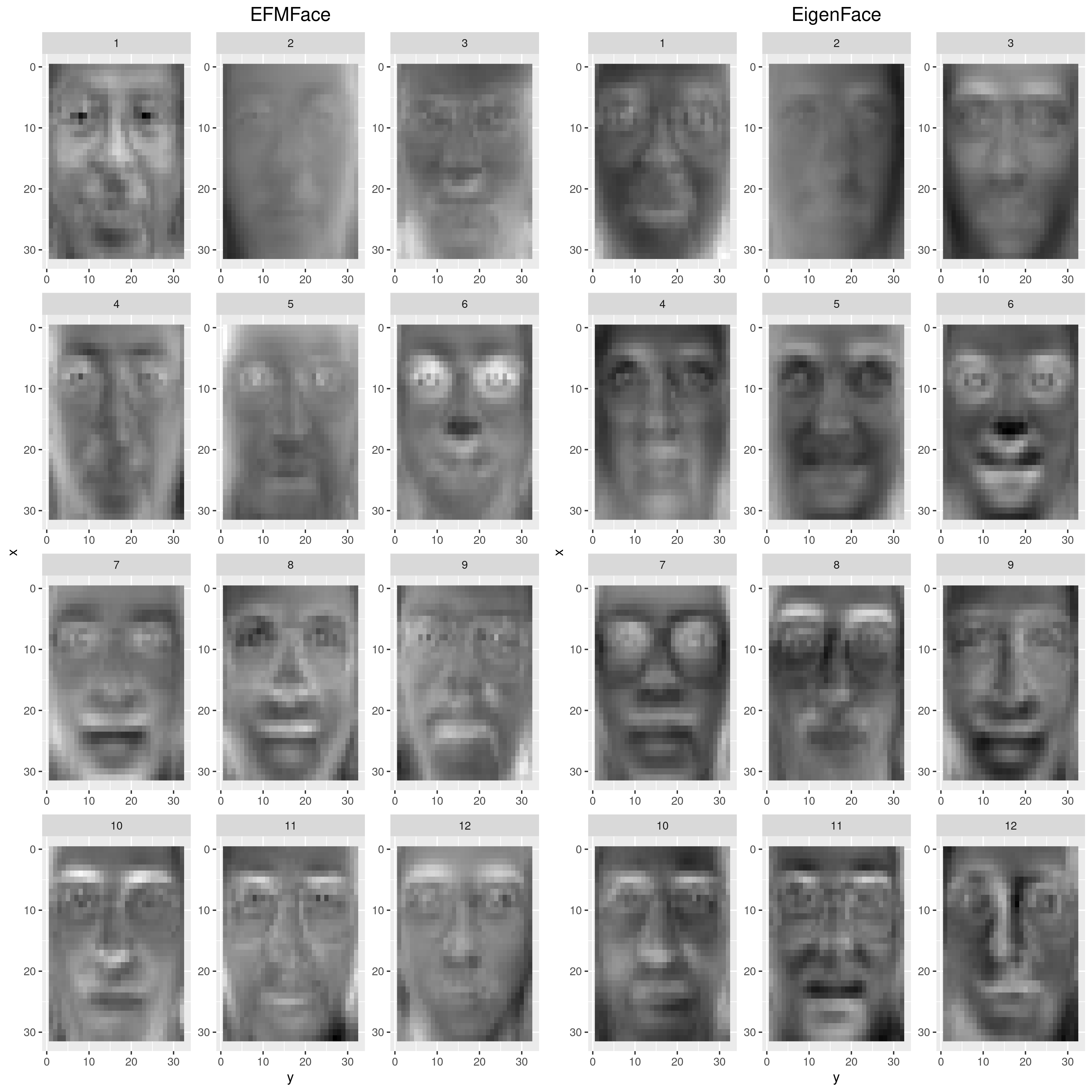}
\caption{Comparison of first 12 latent face loadings from EFM
(left 4-by-3 panel) and EigenFace (right panel).}
\label{fig:eigenface}
\end{figure}

Next, we estimate the factors $\hat{\Lambda}$ using the regularized GLM
regression in~\eqref{eq:recoverfactor}. This extra step is not needed for
Eigenface because both $\Lambda$ and $V$ are estimated jointly. The first 12
latent faces can be visualized in Figure~\ref{fig:eigenface}, along with the
latent space from Eigenface.
After visually comparing those learned faces' vectors, we can conclude that
the EigenFace vectors appear to have more noise and sharper edges in some
regions, indicating that it may capture finer details but also more variations
that might not be as useful for classification. The EFM face vector appears
smoother, suggesting that it may filter out some noise and emphasize more
structured facial features.

\begin{figure}[H]
\centering
\includegraphics[width=.7\textwidth]{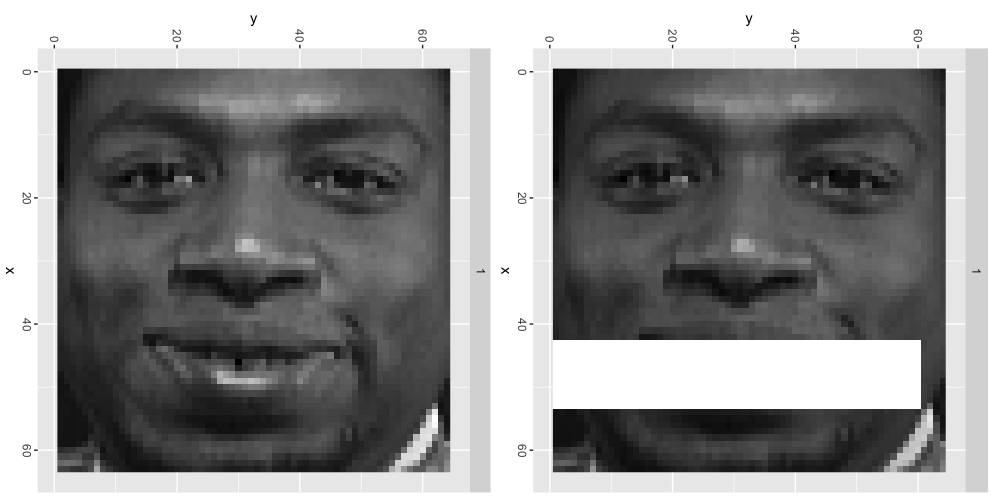}
\includegraphics[width=.7\textwidth]{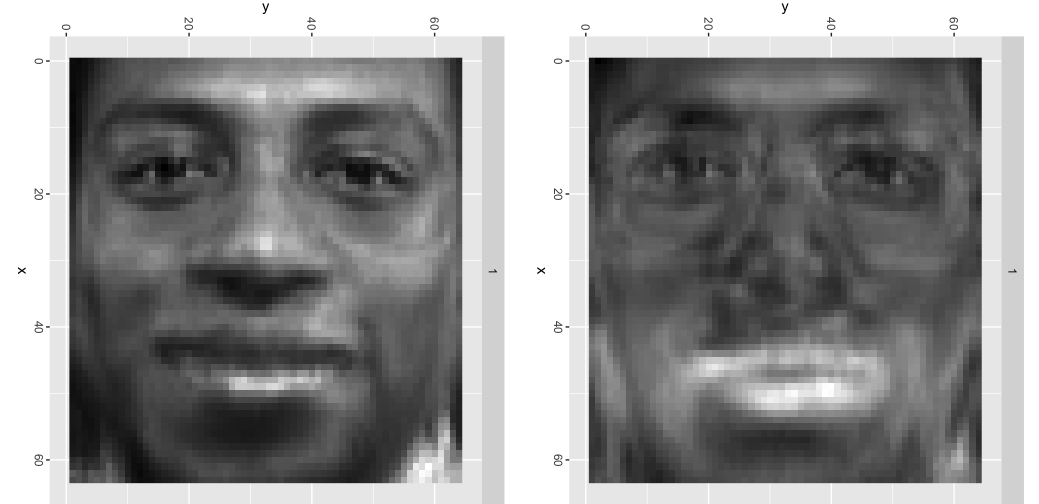}
\caption{Example of reconstruction of cropped face (top row) using EFM and
Eigen-Face (bottom row, left and right, respectively).}
\label{fig:crop_face}
\end{figure}

Finally, we show how the EFM can be used for prediction from partially
observed data. To this end, we select a random face $x^*$ and crop out a
rectangular region of it, in effect partitioning $x^*$ into an observed set
$x^*_{\text{obs}}$ and a missing part $x^*_{\text{miss}}$, as depicted in the
top row of Figure~\ref{fig:crop_face}.

Since for any pixel $i \in [n]$ we have $x^*_i \,|\, \Lambda, v^*
\stackrel{\text{ind}}{\sim} F(g^{-1}(\Lambda_i^\top v^*))$ as
per~\eqref{eq:generate} with the pixel factors $\Lambda$ and new loading
$v^*$, we first fix our estimate $\hat{\Lambda}$ from the training image set
as above and then proceed in two steps:
\begin{enumerate}[(i)]
\item First infer $\hat{v}^*$ by GLM regressing $x^*_{\text{obs}}$ on
  $\hat{\Lambda}$, that is,
\[
\hat{v}^* = \argmax_{v^* \in \mathbb{R}^q} \sum_{i \in \text{obs}}
\log f_{v^*, \hat{\eta}_0, \hat{\Phi}}(x^*_i \,|\, \hat{\Lambda}_i, v^*),
\]
where $\hat{\eta}_0$ and $\hat{\Phi}$ were estimated in the EFM;

\item Next, estimate the missing entries using their means:
$\hat{x}^*_i = g^{-1}(\hat{\Lambda}_i^\top \hat{v}^*)$ for $i \in
\text{miss}$.
\end{enumerate}

In the bottom row of Figure~\ref{fig:crop_face} we see the results from the
reconstruction procedure for a negative binomial EFM and Eigen-Face. To better
appreciate the smaller amount of variability in the estimates from the EFM, we
plot $\hat{x}^*_i$ for all pixels, not just the missing ones. As we can see,
the EFM provides an overall sharper reconstruction and better predictions in
the cropped region.

\subsection{Network analysis}
Another interesting application of our weighted EFM is on social network
analysis, where interest often lies in summarizing a large adjacency matrix
using lower-dimensional representations named ``node embeddings''.
Recently, there has been growing interest in node embedding inference based on
multiple networks. These multiple networks are formally observed as multiple
interaction graphs, $\{\mathcal{G}^{(1)}, \ldots, \mathcal{G}^{(k)}\}$, which
consist of multiple edge relationship, $\{\mathcal{E}^{(1)}, \ldots,
\mathcal{E}^{(k)}\}$, for the same sets of vertices $V$. This emerging field
of research is named as multi-layer or multiplex network
analysis~\citep{kivela2014multilayer}.
Denoting the number of vertices as $n = |V|$, multiplex network inference
starts by transforming these multiple graphs into adjacency matrices
$\{A^{(1)}, \ldots, A^{(k)}\}$ of same dimension $n\times n$.
Factorization and joint inference on these constructed $\{A^{(1)}, \ldots,
A^{(k)}\}$ have been shown to provide better node embeddings with potential
applications to community detection and link
prediction~\citep{wang2019joint,jones2020multilayer}.

One method to enable the joint inference on adjacency matrices is to combine
$\{A^{(1)}, \ldots, A^{(k)}\}$ into a single adjacency matrix $\tilde{A}$.
\citet[][Chapter~2]{draves2022joint} provides a comprehensive introduction to
various aggregation techniques. We briefly summarize some of the relevant
aggregation techniques used in this section:
\begin{description}
    \item[Average Adjacency Spectral Embedding]
      \citep[AASE;][]{tang2018connectome}, which simply averages the adjacency
      matrix through
    \begin{equation}
      \tilde{A}_{ij} = \frac{1}{k}\sum_{l=1}^k A_{ij}^{(l)}
    \end{equation}
    \item[Unfolded Adjacency Spectral Embedding] (UASE), that concatenates the
      adjacency matrices column-wise,
    \begin{equation}
      \tilde{A} = [A^{(1)} \, A^{(2)} \, \cdots \, A^{(k)}] 
    \end{equation}

    \item[Omnibus Embedding] (OE), by constructing pair-wisely the following
      adjacency matrix:
\begin{equation}
\tilde{A} =
\begin{bmatrix}
A^{(1)}  &\frac{1}{2} (A^{(1)} + A^{(2)})  &\cdots  &\frac{1}{2} (A^{(1)} + A^{(k)}) \\
\frac{1}{2} (A^{(1)} + A^{(2)})  & A^{(2)}  &\cdots  &\frac{1}{2} (A^{(2)} + A^{(k)}) \\
\vdots &\vdots  &\ddots  &\vdots \\
\frac{1}{2} (A^{(k)} + A^{(1)})  & \frac{1}{2} (A^{(k)} + A^{(2)})  &\cdots  &A^{(k)}
\end{bmatrix}
\end{equation}
\end{description}

Thus, for a given decomposition rank $q$ we can compute a truncated SVD for
the asymmetric $\tilde{A}$ from UASE and a truncated eigendecomposition for
the symmetric $\tilde{A}$ for ASE and OE,
\begin{equation}
\begin{aligned}
    \tilde{A}^{(\text{ASE})} &= UDU^\top, U \in \mathcal{S}_{n,q}(\mathbb{R}),\\
    \tilde{A}^{(\text{UASE})} &= UDV^\top, U \in \mathcal{S}_{n,q}(\mathbb{R}), V\in \mathcal{S}_{nk,q}(\mathbb{R}),\\
    \tilde{A}^{(\text{OE})}&= UDU^\top, U \in \mathcal{S}_{nk,q}(\mathbb{R}).
    \label{eq:agg_embedding}
\end{aligned}
\end{equation}
where $D =\text{Diag}(d_1, \ldots, d_q), d_1\geq\ldots \geq d_q > 0$.
The node embeddings $\Lambda$ can then be defined as
$\Lambda = U_{1:n}D^{1/2}$ with dimension $n\times q$, where $U_{1:n}$ are the
first $n$ rows of the eigenvectors $U$ for the Omnibus
embedding~\citep{levin2017central}.

One significant limitation of these aggregation approaches is that
factorization implicitly treats each layer as contributing equally. In
reality, each network layer can differ greatly, particularly in terms of
sparsity. Equal weighting risks over-emphasizing interactions from dense
graphs while underrepresenting those from sparse graphs. Furthermore, in
network time series---where edge relationships between the same vertices are
tracked over time---it is more sensible to assign greater weight to recent
adjacency matrices than to aggregate all matrices equally, since recent
interactions are likely more informative for predicting future network
behavior.

A straightforward improvement over equal aggregation of individual networks is
to assign different weights during factor inference. EFM facilitates this
approach by supporting both entry-wise and layer-wise weighting of aggregated
interactions. This flexibility in specifying heuristic weights enables us to
obtain improved embeddings for multiplex network analysis.

Specifically, with $\lambda_{\text{max}}^{(k)}$ denoted as the largest
eigenvalue of adjacency matrix $A^{(k)}$, we
\begin{itemize}
    \item set a zero weight to diagonal entries since they carry no
      information about interactions;
    \item uniformly weight all interactions in the $k$-th network by
      $1/\lambda_{\text{max}}^{(k)}$ to ensure that each matrix has its
      largest eigenvalue set to 1;
    \item uniformly weight all the off-diagonal zero interactions as the
      minimal value of the non-zero terms in the weight matrix to reduce the
      bias effect towards zero of no interactions across all networks.
\end{itemize}
The weights are then defined as:
\begin{equation}
\begin{aligned}
     W_{ij} &= \begin{cases}
      0, & \text{if~} i = j, \\
    \sum_{l=1}^k \frac{1}{\lambda^{(l)}_{\text{max}}} A_{ij}^{(l)}, & \text{if~} i \neq j \text{~and~} \sum_{l=1}^k A_{ij}^{l} > 0,\\
    \min_{i, j =1,\ldots,n} \Big\{\sum_{l=1}^k \frac{1}{\lambda^{(l)}_{\text{max}}} A_{ij}^{(l)} \,:\, \sum_{l=1}^k A_{ij}^{(l)} > 0 \Big\}, & \text{if~} i \neq j \text{~and~} \sum_{l=1}^k A_{ij}^{(l)} = 0.\\
    \end{cases}
\end{aligned}
\label{eq:weightAUCS}
\end{equation}
Finally, for the aggregated adjacency matrix we assign value 1 to $A_{ij}$
whenever $i$ and $j$ interact in at least one layer, that is, $A_{ij} =
\max_{k} A_{ij}^{(k)}$.

\subsubsection{AUCS dataset}
\begin{figure}[H]
\centering
\includegraphics[width = \textwidth]{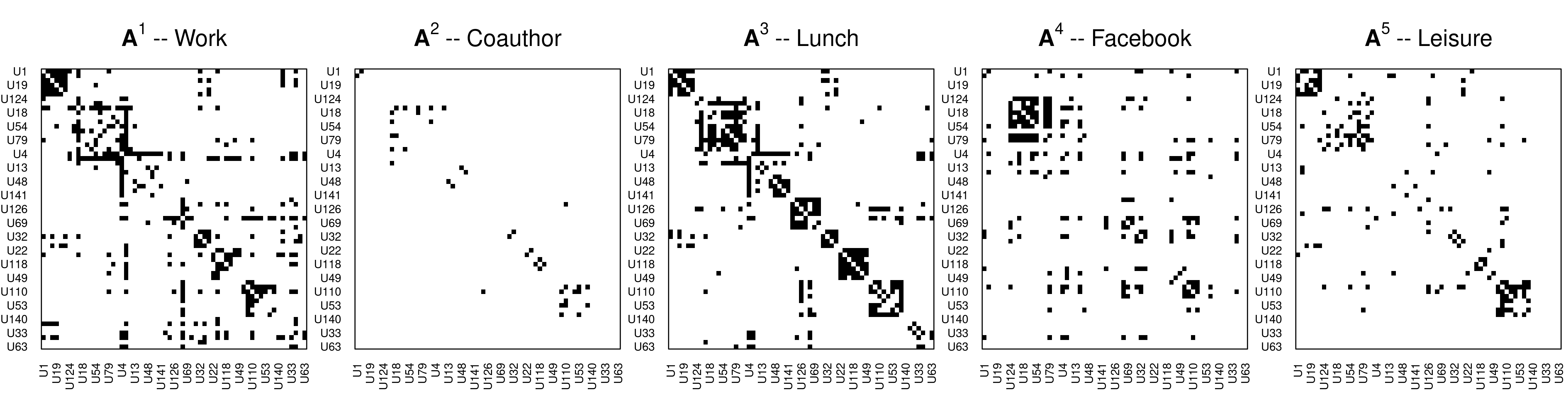}
\caption{Adjacency matrices for each AUCS dataset network. Left to right: work, coauthor, lunch, facebook, and leisure.}
\label{fig:aucs_sparsity}
\end{figure}

We applied our EFM inference to the AUCS
dataset~\citep{dickison2016multilayer}, which records interactions among
$n = 61$ employees in the Department of Computer Science at Aarhus University
across five offline and online contexts: work, coauthor, lunch, Facebook, and
leisure. The sparsity patterns of these interactions are shown in
Figure~\ref{fig:aucs_sparsity}. As confirmed by the dataset's author, 55 of
these employees belong to one of eight research groups, while 6 are
unaffiliated. The research group assignments for the 55 employees serve as
ground truth to evaluate community detection.

Examining the interaction networks reveals that the coauthor layer is notably
sparser than the others. Following the heuristic introduced earlier, we
propose weighting interactions from each network layer according to its
sparsity, in order to improve community inference effectiveness.

\begin{figure}[h]
\centering
\includegraphics[width=0.24\textwidth,height =1.8in]{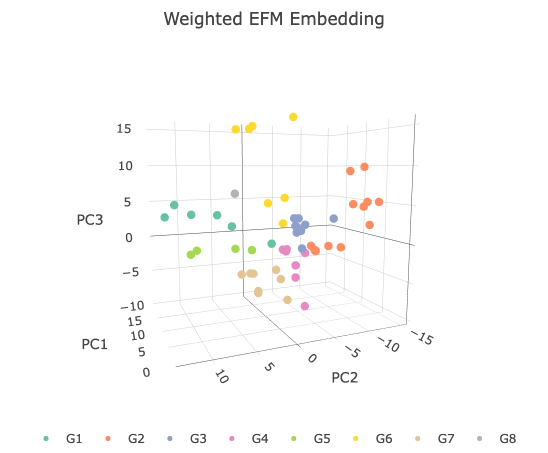}
\includegraphics[width=0.25\textwidth,height =1.8in]{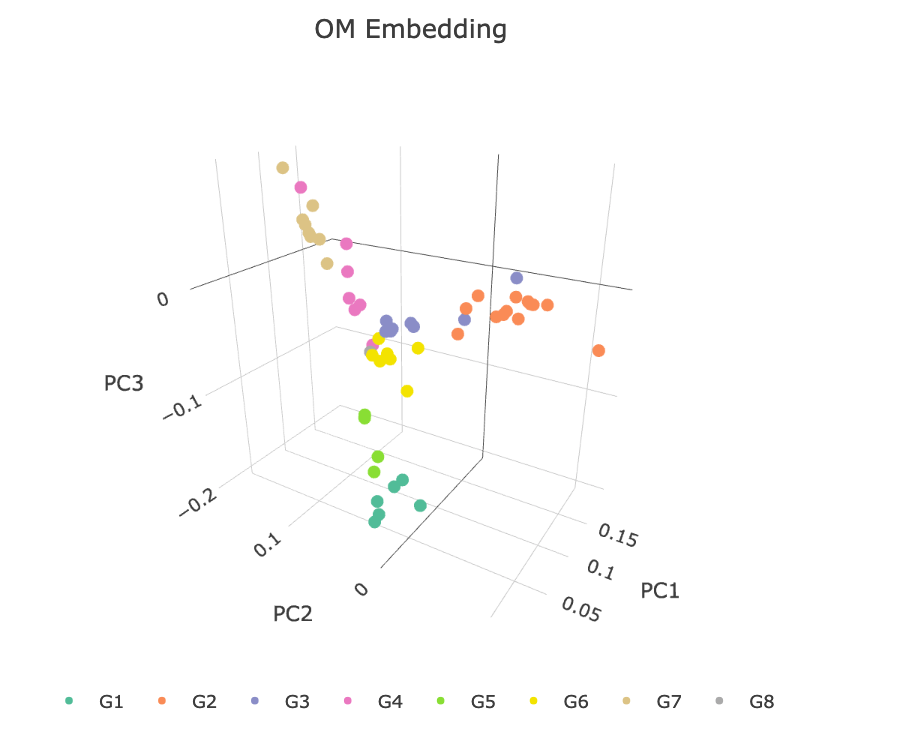}
\includegraphics[width=0.25\textwidth,height =1.8in]{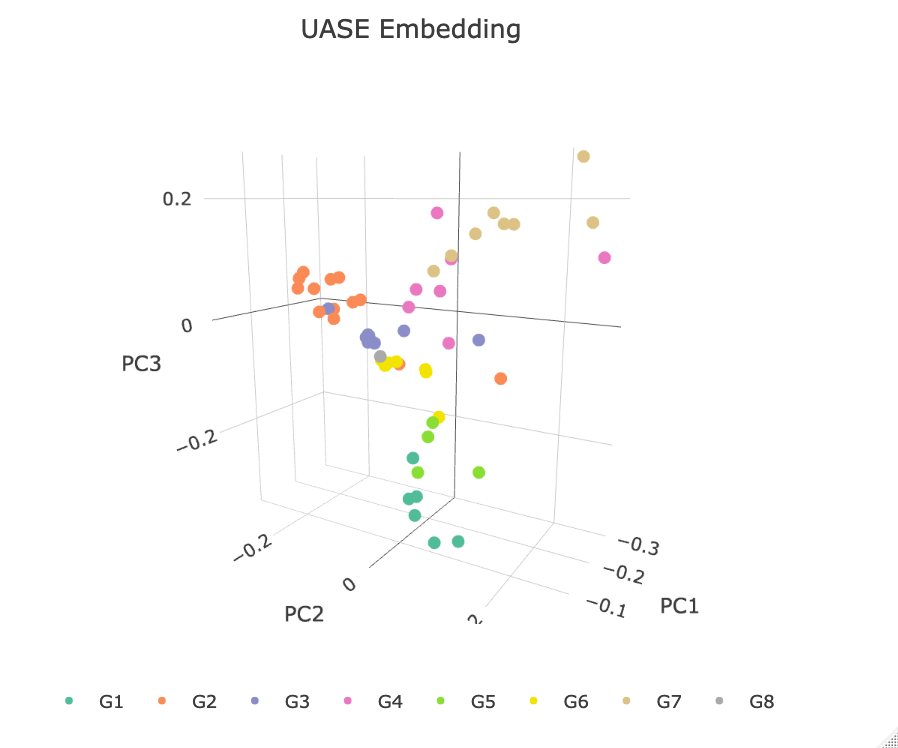}
\includegraphics[width=0.24\textwidth,height =1.8in]{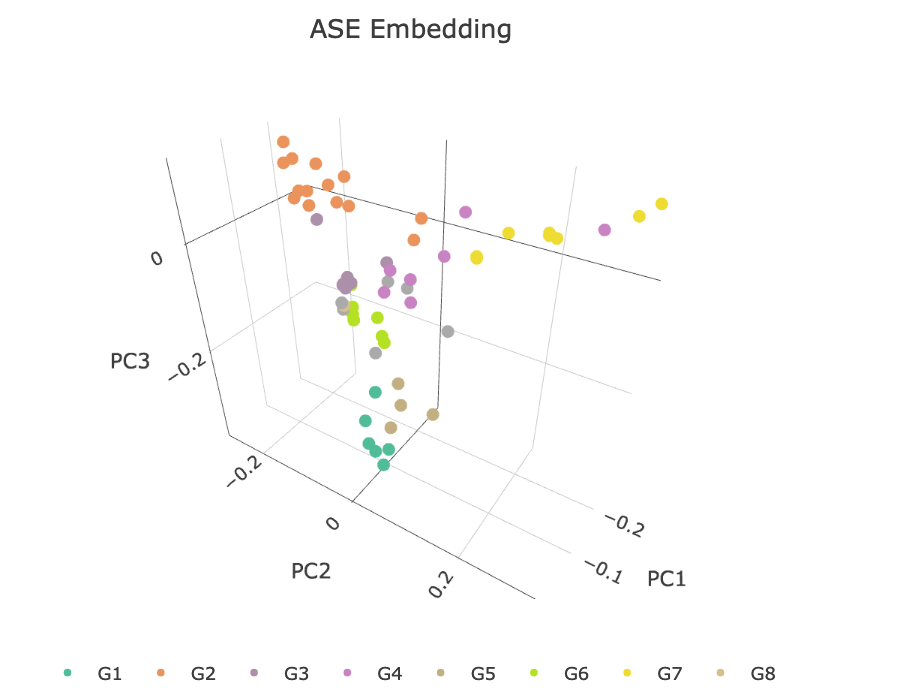}
\caption{Comparing rank $q = 3$ node embeddings for the AUCS dataset. Left to
right: weighted EFM factorization, OM eigendecomposition, UASE SVD
decomposition, and ASE eigendecomposition. Points are colored according to the
research group to which the corresponding employee belongs.}
\label{fig:AUCS_EmbedVis}
\end{figure}

With weights $W$ defined as in~\eqref{eq:weightAUCS}, we then apply a binomial
EFM with logit link to obtain three dimensional ($q=3$) node embeddings of
$\tilde{A}^{\text{(UASE)}}$. We can then visualize those factorized embedding
according the separability of the known research group community labels. For a
comparison to the $\{\text{AASE, UASE, OM}\}$ embedding techniques, SVD or
eigen-decomposition are also conducted on their corresponding aggregated
adjacency matrices~$\tilde{A}$ to obtain their corresponding node embeddings
defined in Eq~\eqref{eq:agg_embedding}.
A visualization with the factorized embeddings $\Lambda = UD^{1/2}$ is shown
in Figure~\ref{fig:AUCS_EmbedVis}.
As we can see, the weighted EFM better separated the research groups as
compared to a naive SVD on any of the embedding approaches, with a more
compact representation of employees within the same research group.
This classification result is also consistent with the existing
literature~\citep{magnani2021analysis} that claims that there are at least
five clear major research groups identified by the publisher of the dataset.

\section{Conclusion}
\label{sec:conclusion}
We introduce two flexible, robust, and efficient classes of optimization
algorithms for exponential family factor modeling. Our approach begins with a
formal treatment of identifiability issues, effectively generalizing the
singular value decomposition to deviance losses. This generalization enables
more representative results across a broader variety of real-world datasets.
While this increased representativity introduces more complexity and greater
computational demands, it often produces lower-rank decompositions compared to
standard SVD, leading to better interpretability and improved predictive
accuracy.

To overcome the main computational challenge---integrating over latent
factors---we propose robust approximation strategies using EM and SGD
optimization. Our algorithms enhance simulated likelihood estimation (SML) by
eliminating asymptotic bias even with moderate simulation sample sizes.
Although EM is deterministic and typically more precise, it is computationally
intensive and best suited for low-rank decompositions (as demonstrated
by~\cite{moustaki2000generalized}, who limit their approach to rank 2).
For higher-rank problems, we implement a more scalable alternative based on
SGD. Leveraging SGD, our EFM approach generalizes better than competing
methods such as DMF. Both our simulation and empirical results consistently
demonstrate these advantages. To encourage further practical adoption, we
provide an efficient R package implementing our proposed algorithms.

Looking ahead, we plan to develop Markov chain Monte Carlo methods to sample
from the posterior distributions of loadings ($V$) and variances ($\Phi$),
allowing us to better characterize inferential uncertainty. While we can
ensure the identifiability of $V$ after model optimization, posterior sampling
requires preserving identifiability at each iteration, necessitating more
sophisticated and specialized algorithms. Another promising future direction
is extending our framework to tensor factor analysis, such as applying
multinomial factor models for categorical or ordinal data.

\bibliographystyle{agsm}
\bibliography{main}

@article{1998ppca,
  author = "Michael E. Tipping and Christopher M. Bishop",
  title = "Probabilistic principal component analysis",
  journal = "Journal of the Royal Statistical Society, Series B ",
  year = 1998
}

@article{2015hpf,
  author = "Prem Gopalan and Jake M. Hofman and David M. Blei",
  title = "Scalable Recommendation with Hierarchical Poisson Factorization",
  journal = "Proceedings of the Thirty-First Conference on Uncertainty in Artificial Intelligence",
  year = 2015
}

@article{1999nmf,
  author = "Daniel D. Lee and Seung Hoe Choi",
  title = "Learning the parts of objects by nonnegative matrix factorization",
  journal = "Nature",
  volume = 401,
  year = 1999
}

@article{2020frank,
  author = "Jianqing Fan and Jianhua Guo and Shurong Zheng",
  title = "Estimating Number of Factors by Adjusted Eigenvalues Thresholding",
  journal = "Journal of the American Statistical Association",
  year = 2020
}

@article{1999convex,
  author = "Dimitri P. Bertsekas",
  title = "Nonlinear Programming: 2nd Edition",
  journal = "Athena Scientific",
  year = 1999
}

@article{2015adam,
  author = "Diederik P. Kingma and Jimmy Ba",
  title = "Adam: A Method for Stochastic Optimization",
  journal = "3rd International Conference on Learning Representations, ICLR 2015",
  year = 2015
}

@article{2011adagrad,
  author = "Duchi, John and Hazan, Elad and Singer, Yoram",
  title = "Adaptive subgradient methods for online learning and stochastic optimization",
  journal = "The Journal of Machine Learning Research",
  year = 2011
}

@article{2009inla,
  author = "Rue, Havard and Sara Martino and Nicolas Chopin",
  title = "Approximate Bayesian inference for latent Gaussian models by using integrated nested Laplace approximations",
  journal = "Journal of the Royal Statistical Society, Series B",
  year = 2009
}

@article{2012rmsprop,
  author = "Ieleman, T. and Hinton, G",
  title = "Neural Networks for Machine Learning",
  journal = "Technical report",
  year = 2012
}

@book{1994factor,
  author = "Alexander Basilevsky",
  title = "Statistical Factor Analysis and Related Methods",
  publisher = "John Wiley \& Sons",
  year = 1994
}

@article{1963factor,
  author = "Theodore Wilbur Anderson",
  title = "The Use of Factor Analysis in the Statistical Analysis of Multiple Time Series",
  journal = "Psychometrika",
  year = 1963
}

@article{2015famma,
  author = "Fama, Eugene F and Kenneth R. French",
  title = "A Five-Factor Asset Pricing Model",
  journal = "Journal of Financial Economics",
  year = 2015
}

@article{2021biofactor,
  author = "Tianchen Xu and Ryan T. and Demmer Gen Li",
  title = "Zero-Inflated Poisson Factor Model with Application to Microbiome Read Counts",
  journal = "Biometrics",
  year = 2021
}

@article{1986psyfactor,
  author = "Ford, J. Kevin and MacCallum, Robert C. and Tait, Marianne",
  title = "The Application of Exploratory Factor Analysis in Applied Psychology: A Critical Review and Analysis",
  journal = "Personnel Psychology",
  year = 1986
}

@article{2003adfactor,
  author = "Ningning Wu and Jing Zhang",
  title = "Factor Analysis Based Anomaly Detection",
  journal = "IEEE Systems, Man and Cybernetics SocietyInformation Assurance Workshop",
  year = 2003
}

@article{2008facefactor,
  author = "Simon J D Prince and James H Elder and Jonathan Warrell and Fatima M Felisberti",
  title = "Tied Factor Analysis for Face Recognition across Large Pose Differences",
  journal = "IEEE Transactions on Pattern Analysis and Machine Intelligence",
  year = 2008
}

@article{2003expfactor,
  author = "Michel Wedel and Ulf Bo Ckenholt and Wagner A. Kamakurac",
  title = "Factor Models for Multivariate Count Data",
  journal = "Journal of Multivariate Analysis",
  year = 2003
}

@article{2006emadam,
  author = "Jank, Wolfgang",
  title = "Implementing and Diagnosing the Stochastic Approximation EM Algorithm",
  journal = "Journal of Computational and Graphical Statistics",
  volume = 15,
  pages = {803–29},
  year = 2006
}

@article{2001expfactor2,
  author = "Wedel, Michel and Wagner A. Kamakura",
  title = "Factor Analysis with (Mixed) Observed and Latent Variables in the Exponential Family",
  journal = "Psychometrika",
  volume = 66,
  pages = {513-30},
  year = 2001
}

@article{2003MCEMcomp,
  author = "Jank, Wolfgang and James Booth",
  title = "Efficiency of Monte Carlo EM and Simulated Maximum Likelihood in Two-Stage Hierarchical Models",
  journal = "Journal of Computational and Graphical Statistics",
  volume = 12,
  pages = {214-29},
  year = 2003
}

@article{1995smlsize,
  author = "Lee, Lung-fei",
  title = "Asymptotic Bias in Simulated Maximum Likelihood Estimation of Discrete Choice Models",
  journal = "Econometric Theory",
  volume = 11,
  pages = {437-83},
  year = 1995
}

@article{1986accurate,
  title={Accurate approximations for posterior moments and marginal densities},
  author={Tierney, Luke and Kadane, Joseph B},
  journal={Journal of the american statistical association},
  volume={81},
  number={393},
  pages={82--86},
  year={1986},
  publisher={Taylor \& Francis}
}

@article{1989fully,
  title={Fully exponential Laplace approximations to expectations and variances of nonpositive functions},
  author={Tierney, Luke and Kass, Robert E and Kadane, Joseph B},
  journal={Journal of the american statistical association},
  volume={84},
  number={407},
  pages={710--716},
  year={1989},
  publisher={Taylor \& Francis}
}

@article{2021dmf,
  title={Deviance matrix factorization},
  author={Wang, Liang and Carvalho, Luis},
  journal={Electronic Journal of Statistics},
  volume={17},
  number={2},
  pages={3762--3810},
  year={2023},
  publisher={The Institute of Mathematical Statistics and the Bernoulli Society}
}

@misc{2017mnist,
  author       = {Han Xiao and Kashif Rasul and Roland Vollgraf},
  title        = {Fashion-MNIST: a Novel Image Dataset for Benchmarking Machine Learning Algorithms},
  date         = {2017-08-28},
  year         = {2017},
  eprintclass  = {cs.LG},
  eprinttype   = {arXiv},
  eprint       = {cs.LG/1708.07747},
}

@article{hoff2002latent,
  title={Latent space approaches to social network analysis},
  author={Hoff, Peter D and Raftery, Adrian E and Handcock, Mark S},
  journal={Journal of the american Statistical association},
  volume={97},
  number={460},
  pages={1090--1098},
  year={2002},
  publisher={Taylor \& Francis}
}

@inproceedings{young2007random,
  title={Random dot product graph models for social networks},
  author={Young, Stephen J and Scheinerman, Edward R},
  booktitle={Algorithms and Models for the Web-Graph: 5th International Workshop, WAW 2007, San Diego, CA, USA, December 11-12, 2007. Proceedings 5},
  pages={138--149},
  year={2007},
  organization={Springer}
}

@article{levy2014neural,
  title={Neural word embedding as implicit matrix factorization},
  author={Levy, Omer and Goldberg, Yoav},
  journal={Advances in neural information processing systems},
  volume={27},
  year={2014}
}

@article{mikolov2013efficient,
  title={Efficient estimation of word representations in vector space},
  author={Mikolov, Tomas and Chen, Kai and Corrado, Greg and Dean, Jeffrey},
  journal={arXiv preprint arXiv:1301.3781},
  year={2013}
}

@article{shapiro1985identifiability,
  title={Identifiability of factor analysis: Some results and open problems},
  author={Shapiro, Alexander},
  journal={Linear Algebra and its Applications},
  volume={70},
  pages={1--7},
  year={1985},
  publisher={Elsevier}
}

@article{borenstein2010basic,
  title={A basic introduction to fixed-effect and random-effects models for meta-analysis},
  author={Borenstein, Michael and Hedges, Larry V and Higgins, Julian PT and Rothstein, Hannah R},
  journal={Research synthesis methods},
  volume={1},
  number={2},
  pages={97--111},
  year={2010},
  publisher={Wiley Online Library}
}

@article{kitagawa1987non,
  title={Non-gaussian state—space modeling of nonstationary time series},
  author={Kitagawa, Genshiro},
  journal={Journal of the American statistical association},
  volume={82},
  number={400},
  pages={1032--1041},
  year={1987},
  publisher={Taylor \& Francis}
}

@article{gribonval2010dictionary,
  title={Dictionary Identification—Sparse Matrix-Factorization via $l\_1 $-Minimization},
  author={Gribonval, R{\'e}mi and Schnass, Karin},
  journal={IEEE Transactions on Information Theory},
  volume={56},
  number={7},
  pages={3523--3539},
  year={2010},
  publisher={IEEE}
}

@article{li2010nonnegative,
  title={Nonnegative matrix factorization on orthogonal subspace},
  author={Li, Zhao and Wu, Xindong and Peng, Hong},
  journal={Pattern Recognition Letters},
  volume={31},
  number={9},
  pages={905--911},
  year={2010},
  publisher={Elsevier}
}

@book{ma2012manifold,
  title={Manifold learning theory and applications},
  author={Ma, Yunqian and Fu, Yun},
  volume={434},
  year={2012},
  publisher={CRC press Boca Raton}
}

@article{carter1996markov,
  title={Markov chain Monte Carlo in conditionally Gaussian state space models},
  author={Carter, Chris K and Kohn, Robert},
  journal={Biometrika},
  volume={83},
  number={3},
  pages={589--601},
  year={1996},
  publisher={Oxford University Press}
}

@article{2014glove,
  author = "Jeffrey Pennington, Richard Socher and Christopher Manning",
  title = "GloVe: Global Vectors for Word Representation",
  journal = "Proceedings of the 2014 Conference on Empirical Methods in Natural Language Processing (EMNLP)",
    pages={1532--1543},
  year = 2014}

@inproceedings{kalayeh2014nmf,
  title={NMF-KNN: Image annotation using weighted multi-view non-negative matrix factorization},
  author={Kalayeh, Mahdi M and Idrees, Haroon and Shah, Mubarak},
  booktitle={Proceedings of the IEEE conference on computer vision and pattern recognition},
  pages={184--191},
  year={2014}
}

@article{reimann2002factor,
  title={Factor analysis applied to regional geochemical data: problems and possibilities},
  author={Reimann, Clemens and Filzmoser, Peter and Garrett, Robert G},
  journal={Applied geochemistry},
  volume={17},
  number={3},
  pages={185--206},
  year={2002},
  publisher={Elsevier}
}

@article{caffo2005ascent,
  title={Ascent-based Monte Carlo expectation--maximization},
  author={Caffo, Brian S and Jank, Wolfgang and Jones, Galin L},
  journal={Journal of the Royal Statistical Society: Series B (Statistical Methodology)},
  volume={67},
  number={2},
  pages={235--251},
  year={2005},
  publisher={Wiley Online Library}
}

@article{davenport20141,
  title={1-bit matrix completion},
  author={Davenport, Mark A and Plan, Yaniv and Van Den Berg, Ewout and Wootters, Mary},
  journal={Information and Inference: A Journal of the IMA},
  volume={3},
  number={3},
  pages={189--223},
  year={2014},
  publisher={OUP}
}

@article{kivela2014multilayer,
  title={Multilayer networks},
  author={Kivel{\"a}, Mikko and Arenas, Alex and Barthelemy, Marc and Gleeson, James P and Moreno, Yamir and Porter, Mason A},
  journal={Journal of complex networks},
  volume={2},
  number={3},
  pages={203--271},
  year={2014},
  publisher={Oxford University Press}
}

@article{levin2017central,
  title={A central limit theorem for an omnibus embedding of multiple random graphs and implications for multiscale network inference},
  author={Levin, Keith and Athreya, Avanti and Tang, Minh and Lyzinski, Vince and Park, Youngser and Priebe, Carey E},
  journal={arXiv preprint arXiv:1705.09355},
  year={2017}
}

@article{wang2019joint,
  title={Joint embedding of graphs},
  author={Wang, Shangsi and Arroyo, Jes{\'u}s and Vogelstein, Joshua T and Priebe, Carey E},
  journal={IEEE transactions on pattern analysis and machine intelligence},
  volume={43},
  number={4},
  pages={1324--1336},
  year={2019},
  publisher={IEEE}
}

@article{jones2020multilayer,
  title={The multilayer random dot product graph},
  author={Jones, Andrew and Rubin-Delanchy, Patrick},
  journal={arXiv preprint arXiv:2007.10455},
  year={2020}
}

@book{dickison2016multilayer,
  title={Multilayer social networks},
  author={Dickison, Mark E and Magnani, Matteo and Rossi, Luca},
  year={2016},
  publisher={Cambridge University Press}
}

@phdthesis{draves2022joint,
  title={Joint spectral embeddings of random dot product graphs},
  author={Draves, Benjamin},
  school ={Boston University},
  year={2022}
}

@article{tang2018connectome,
  title={Connectome smoothing via low-rank approximations},
  author={Tang, Runze and Ketcha, Michael and Badea, Alexandra and Calabrese, Evan D and Margulies, Daniel S and Vogelstein, Joshua T and Priebe, Carey E and Sussman, Daniel L},
  journal={IEEE transactions on medical imaging},
  volume={38},
  number={6},
  pages={1446--1456},
  year={2018},
  publisher={IEEE}
}

@article{magnani2021analysis,
  title={Analysis of multiplex social networks with R},
  author={Magnani, Matteo and Rossi, Luca and Vega, Davide},
  journal={Journal of Statistical Software},
  volume={98},
  pages={1--30},
  year={2021}
}

@article{fan2008high,
  title={High dimensional covariance matrix estimation using a factor model},
  author={Fan, Jianqing and Fan, Yingying and Lv, Jinchi},
  journal={Journal of Econometrics},
  volume={147},
  number={1},
  pages={186--197},
  year={2008},
  publisher={Elsevier}
}

@article{nelder1987extended,
  title={An extended quasi-likelihood function},
  author={Nelder, John A and Pregibon, Daryl},
  journal={Biometrika},
  volume={74},
  number={2},
  pages={221--232},
  year={1987},
  publisher={Oxford University Press}
}

@book{bartholomew2011,
  title={Latent Variable Models and Factor Analysis: a Unified Approach},
  author={Bartholomew, David J and Knott, Martin and Moustaki, Irini},
  year={2011},
  publisher={John Wiley \& Sons}
}

@article{kaiser1958varimax,
  title={The varimax criterion for analytic rotation in factor analysis},
  author={Kaiser, Henry F},
  journal={Psychometrika},
  volume={23},
  number={3},
  pages={187--200},
  year={1958},
  publisher={Springer}
}

@article{xia2020average,
  title={The average of a negative-binomial L{\'e}vy process and a class of Lerch distributions},
  author={Xia, Weixuan},
  journal={Communications in Statistics-Theory and Methods},
  volume={49},
  number={4},
  pages={1008--1024},
  year={2020},
  publisher={Taylor \& Francis}
}

@article{li2024comprehensive,
  title={Comprehensive evaluation of Mal-API-2019 dataset by machine learning in malware detection},
  author={Li, Zhenglin and Zhu, Haibei and Liu, Houze and Song, Jintong and Cheng, Qishuo},
  journal={arXiv preprint arXiv:2403.02232},
  year={2024}
}

@article{wedderburn1974quasi,
  title={Quasi-likelihood functions, generalized linear models, and the {G}auss-{N}ewton method},
  author={Wedderburn, Robert WM},
  journal={Biometrika},
  volume={61},
  number={3},
  pages={439--447},
  year={1974},
  publisher={Oxford University Press}
}

@article{dempster1977em,
  title={Maximum likelihood from incomplete data via the {EM} algorithm},
  author={Dempster, Arthur P and Laird, Nan M and Rubin, Donald B},
  journal={Journal of the Royal Statistical Society: Series B (Methodological)},
  volume={39},
  number={1},
  pages={1--22},
  year={1977},
  publisher={Wiley Online Library}
}

@article{golub1969gauss,
  title={Calculation of {G}auss quadrature rules},
  author={Golub, Gene H and Welsch, John H},
  journal={Mathematics of computation},
  volume={23},
  number={106},
  pages={221--230},
  year={1969}
}

@article{zhu2019multi,
  title={Multi-view deep subspace clustering networks},
  author={Zhu, Pengfei and Hui, Binyuan and Zhang, Changqing and Du, Dawei and Wen, Longyin and Hu, Qinghua},
  journal={arXiv preprint arXiv:1908.01978},
  year={2019}
}

@article{robbins1951stochastic,
  title={A stochastic approximation method},
  author={Robbins, Herbert and Monro, Sutton},
  journal={The annals of mathematical statistics},
  pages={400--407},
  year={1951},
  publisher={JSTOR}
}

@article{computers13060151,
  author = {Ruan, Kangrui and Di, Xuan},
  title = {{InfoSTGCAN}: An Information-Maximizing Spatial-Temporal Graph Convolutional Attention Network for Heterogeneous Human Trajectory Prediction},
  journal = {Computers},
  volume = {13},
  year = {2024},
  number = {6},
  article-number = {151},
  url = {https://www.mdpi.com/2073-431X/13/6/151}
}

@book{mccullaghnelder,
  address = {London},
  author = {McCullagh, P. and Nelder, J. A.},
  date = {(1989)},
  publisher = {Chapman \& Hall / CRC},
  title = {Generalized Linear Models},
  year = 1989
}

@article{hand2001simple,
  title={A simple generalisation of the area under the {ROC} curve for multiple class classification problems},
  author={Hand, David J and Till, Robert J},
  journal={Machine learning},
  volume={45},
  number={2},
  pages={171--186},
  year={2001},
  publisher={Springer}
}

@article{moustaki2000generalized,
  title={Generalized Latent Trait Models},
  author={Moustaki, Irini and Knott, Martin},
  journal={Psychometrika},
  volume={65},
  pages={391--411},
  year={2000},
  publisher={Springer}
}

@article{hinton2002stochastic,
  title={Stochastic neighbor embedding},
  author={Hinton, Geoffrey E and Roweis, Sam},
  journal={Advances in neural information processing systems},
  volume={15},
  year={2002}
}

@article{van2008visualizing,
  title={Visualizing data using {t-SNE}},
  author={Van der Maaten, Laurens and Hinton, Geoffrey},
  journal={Journal of machine learning research},
  volume={9},
  number={11},
  year={2008}
}

@book{breiman2017classification,
  title={Classification and regression trees},
  author={Breiman, Leo and Friedman, Jerome and Olshen, Richard A and Stone, Charles J},
  year={2017},
  publisher={Routledge}
}

@article{cover1967nearest,
  title={Nearest neighbor pattern classification},
  author={Cover, Thomas and Hart, Peter},
  journal={IEEE transactions on information theory},
  volume={13},
  number={1},
  pages={21--27},
  year={1967},
  publisher={IEEE}
}

@inproceedings{turk1991face,
  title={Face Recognition Using Eigenfaces},
  author={Turk, Matthew A and Pentland, Alex},
  booktitle={Proc IEEE Conference on Computer Vision and Pattern Recognition},
  volume={91},
  pages={586--591},
  year={1991}
}

@article{wei1990monte,
  title={A Monte Carlo implementation of the EM algorithm and the poor man's data augmentation algorithms},
  author={Wei, Greg CG and Tanner, Martin A},
  journal={Journal of the American statistical Association},
  volume={85},
  number={411},
  pages={699--704},
  year={1990},
  publisher={Taylor \& Francis}
}

@article{glmnet,
  title = {Regularization Paths for Generalized Linear Models via Coordinate Descent},
  author = {Jerome Friedman and Trevor Hastie and Robert Tibshirani},
  journal = {Journal of Statistical Software},
  year = {2010},
  volume = {33},
  number = {1},
  pages = {1--22},
  doi = {10.18637/jss.v033.i01},
}

@book{knight1999mathematical,
  title={Mathematical Statistics},
  author={Knight, Keith},
  year={1999},
  publisher={CRC Press}
}

@article{lange1995gradient,
  title={A gradient algorithm locally equivalent to the EM algorithm},
  author={Lange, Kenneth},
  journal={Journal of the Royal Statistical Society: Series B (Methodological)},
  volume={57},
  number={2},
  pages={425--437},
  year={1995},
  publisher={Wiley Online Library}
}

@article{lange1995quasi,
  title={A quasi-Newton acceleration of the EM algorithm},
  author={Lange, Kenneth},
  journal={Statistica Sinica},
  pages={1--18},
  year={1995},
  publisher={JSTOR}
}

@book{mclachlan2008algorithm,
  title={The {EM} algorithm and extensions},
  author={McLachlan, Geoffrey J and Krishnan, Thriyambakam},
  year={2008},
  publisher={John Wiley \& Sons}
}

@article{rohe2023vintage,
  title={Vintage factor analysis with {Varimax} performs statistical inference},
  author={Rohe, Karl and Zeng, Muzhe},
  journal={Journal of the Royal Statistical Society Series B: Statistical Methodology},
  volume={85},
  number={4},
  year={2023},
  pages={1037--1060}
}

@article{zhang2022adam,
  title={Adam can converge without any modification on update rules},
  author={Zhang, Yushun and Chen, Congliang and Shi, Naichen and Sun, Ruoyu and Luo, Zhi-Quan},
  journal={Advances in neural information processing systems},
  volume={35},
  pages={28386--28399},
  year={2022}
}

@article{bottou2018optimization,
  title={Optimization methods for large-scale machine learning},
  author={Bottou, L{\'e}on and Curtis, Frank E and Nocedal, Jorge},
  journal={SIAM review},
  volume={60},
  number={2},
  pages={223--311},
  year={2018},
  publisher={SIAM}
}

\end{document}


\section*{Appendix: Fisher scoring and Gaussian posterior approximation}
The derivation of the Fisher scoring updates for quasi-likelihood models is
fairly standard~\citep{mccullaghnelder}, but we list it here for completeness.
As in the main text, to simplify the notation we drop the dependency of
the posterior mode and precision $\hat{\Lambda}_i$ and $H_i$ on data $X_i$ and
current parameter $\theta^{(t)}$ and further denote $\theta^{(t)}$ just as
$\theta$.

We want to find the mode of the posterior $\Lambda_i\,|\,X_i$,
\[
\hat{\Lambda}_i = \argmax_{\Lambda_i} \log f_\theta(X_i, \Lambda_i).
\]
With the log joint at parameter $\theta$ as
$\ell_\theta(\Lambda_i) = \log f_\theta(X_i\,|\,\Lambda_i) +
\log f(\Lambda_i)$, the linear predictors $\eta_i := V\Lambda_i$,
the means $\mu_i = g^{-1}(\eta_i)$ (applied element-wise),
$N_i := \partial \mu_i/\partial \eta_i =
\text{Diag}_{j=1,\ldots,p}\big\{g'(\mu_{ij})^{-1}\big\}$,
and $W_i = \text{Diag}_{j=1,\ldots,p}\big\{\mathcal{V}(\mu_{ij})\big\}$,
\[
\frac{\partial \ell_\theta}{\partial \Lambda_i} :=
U(\Lambda_i) = \frac{1}{\phi}V^\top N_i W_i^{-1} (X_i - \mu_i) - \Lambda_i.
\]
Setting $D_i = N_i V$, we also have
\[
H(\Lambda_i) = -\Exp_{X_i\,|\,\Lambda_i}\Bigg[
\frac{\partial^2 \ell_\theta(\Lambda_i)}%
{\partial\Lambda_i \partial \Lambda_i^\top} \Bigg] =
\frac{1}{\phi} D_i^\top W_i^{-1} D_i + I_q.
\]
The update from Newton's method---here as Fisher scoring since we are using
the expectation of the Hessian, the information---at the $t$-th iteration is
\[
\Lambda_i^{(t+1)} = \Lambda_i^{(t)} + H(\Lambda_i^{(t)})^{-1}
U(\Lambda_i^{(t)}).
\]
Multiplying on the left by $H(\Lambda_i^{(t)})$ yields the iteratively
reweighted least-squares (IRLS) update
\[
H(\Lambda_i^{(t)}) \Lambda_i^{(t+1)} = \frac{1}{\phi}
{D_i^{(t)}}^\top {W_i^{(t)}}^{-1} \big[
D_i^{(t)} \Lambda_i^{(t)} + X_i - \mu_i^{(t)} \big].
\]
The update can be iterated until convergence yielding $\hat{\Lambda}_i$.
This procedure can be done using package \texttt{glmnet}~\citep{glmnet}, for
instance, since a standard normal prior for $\Lambda_i$ is equivalent to a
ridge unit penalty on $\Lambda_i$. We note that using the expected information
$H(\hat{\Lambda}_i)$ results in increased robustness, that is, higher
numerical stability and resistance to outliers~\citep{knight1999mathematical}.

Moreover, a second-order Taylor expansion of $\ell_\theta$ around
$\hat{\Lambda}_i$ yields
\begin{align*}
\ell_\theta(\Lambda_i) & \approx \ell_\theta(\hat{\Lambda}_i) +
(\Lambda_i - \hat{\Lambda}_i)^\top \frac{\partial \ell_\theta}{\partial
\Lambda_i}(\hat{\Lambda}_i) +
\frac{1}{2}(\Lambda_i - \hat{\Lambda}_i)^\top
\frac{\partial \ell_\theta^2}{\partial \Lambda_i \partial \Lambda_i^\top}%
(\hat{\Lambda}_i)
(\Lambda_i - \hat{\Lambda}_i) \\
& \approx
\ell_\theta(\hat{\Lambda}_i) -
\frac{1}{2}(\Lambda_i - \hat{\Lambda}_i)^\top
H(\hat{\Lambda}_i) (\Lambda_i - \hat{\Lambda}_i)
\end{align*}
where in the second approximation the linear term vanishes since $\partial
\ell_\theta/\partial \Lambda_i(\hat{\Lambda}_i) = 0$ and we substitute the
observed information by the expected information $H(\hat{\Lambda}_i)$ as in
Fisher scoring. Thus, with $\hat{H}_i := H(\hat{\Lambda}_i)$,
$\Lambda_i\,|\,X_i \approx N(\hat{\Lambda}_i, \hat{H}_i)$.

\section*{Appendix: Gradient and Hessians for SGD Optimization}
\begin{description}
\item[Gradient for $V_j$:]
\begin{equation}
    \begin{aligned}
        -\nabla_{V_j} \log f_\theta(X_i,\Lambda_i)
        &=   \frac{w_{ij}}{\phi_j} \nabla_{V_j} \bigg(\int_{\mu_{ij}}^{X_{ij}} \frac{1}{\mathcal{V}(t)}(X_{ij} - t) dt  \bigg)
        =  -  \frac{w_{ij}}{\phi_j} \frac{1}{\mathcal{V}(\mu_{ij})}(X_{ij} - \mu_{ij}) \frac{\partial \mu_{ij}}{\partial \eta_{ij}} \frac{\partial \eta_{ij}}{\partial V_j} \\
        & =  -  \frac{w_{ij}}{\phi_j} \frac{1}{\mathcal{V}(\mu_{ij})}(X_{ij} - \mu_{ij})  \frac{1}{g'(\mu_{ij})}\Lambda_i.
    \end{aligned}
\end{equation}

\item[Gradient for $\phi_j$:]
We repeat the derivation for parameter $\phi_j \in \mathbb{R}_+$,
for $j \in [p]$:
\begin{equation}
    \begin{aligned}
        \nabla_{\phi_j} (-\log f_\theta(X_i,\Lambda_i)) &= -\frac{w_{ij}}{\phi_j^2}\bigg(\int_{\mu_{ij}}^{X_{ij}} \frac{1}{\mathcal{V}(t)}(X_i - t) dt  \bigg) + \frac{1}{2\phi_j}
        = - \frac{w_{ij}}{2\phi_j^2} Q(X_{ij} ; \mu_{ij}) + \frac{1}{2\phi_j}\\
        & =  \frac{1}{2\phi_j} \bigg( -\frac{w_{ij}}{\phi_j}Q(X_{ij} ; \mu_{ij}) + 1\bigg)
        \approx \frac{1}{2\phi_j} \bigg(-\frac{w_{ij} (X_{ij} - \mu_{ij})^2}{\phi_j \mathcal{V}(\mu_{ij})} +1  \bigg).
    \end{aligned}
\end{equation}
where $Q(X_{ij} ; \mu_{ij}) =  -2\int_{X_{ij}}^{\mu_{ij}} \frac{1}{\mathcal{V}(t)}(X_i - t) dt$ is the quasi-deviance function.

\item[Gradient for $\eta_0$:]
We repeat the derivation for $\eta_{0j} \in \mathbb{R}$,
for $j \in [p]$:
\begin{equation}
    \begin{aligned}
        \nabla_{\eta_{0j}} (\log f_\theta(X_i,\Lambda_i))
        &=  \frac{w_{ij}}{\phi_j} \nabla_{\eta_{0j}} \bigg(\int_{\mu_{ij}}^{X_{ij}} \frac{1}{\mathcal{V}(t)}(X_i - t) dt  \bigg)
        =  - \frac{w_{ij}}{\phi_j} \frac{1}{\mathcal{V}(\mu_{ij})}(X_{ij} - \mu_{ij}) \frac{\partial \mu_{ij}}{\partial \eta_{ij}} \frac{\partial \eta_{ij}}{\partial \eta_{0j}} \\
        & =  \frac{1}{\Phi_{jj}} \frac{1}{\mathcal{V} (\mu_{ij}) g'(\mu_{ij})} (\mu_{ij} - X_{ij}).
    \end{aligned}
\end{equation}

\item[Hessian for $V_j$:]
In our later optimization, we additionally need the Hessian of the $\log f_\theta(X_i, \Lambda_i)$.
The Hessian for $V_j$ and $\Lambda_i$ are symmetric with respect to each other
with a simple notation change; below we use the former as an example. We need
the following definitions:
\begin{itemize}
    \item $S_{ij} = \frac{w_{ij}}{\phi_j} \frac{{{g^{-1}}'(\eta_{ij})}^2}{V(\mu_{ij})} $ and  $G_{ij} = \frac{{g^{-1}}'(\eta_{ij})}{V(\mu_{ij})} \frac{w_{ij}}{\phi_j} (X_{ij} - \mu_{ij}) $
    \item $D_{i\cdot} = \text{Diag}\{S_{i\cdot}^{(t)}\}$ with $S_{i\cdot}$
denotes the $i$-th row of $S$. Similarly, $\text{Diag}\{S_{\cdot j}^{(t)}\}$ with $S_{\cdot j}$
denotes the $j$-th column of $S$.
\end{itemize}

The Hessian for $V_j$ is then
\begin{multline}
\Exp_{X_i} \bigg[ -\nabla_{V_j}^{2} \log f_\theta(X_i,\Lambda_i)\bigg] =
  \Exp_{X_i} \bigg[\bigg(\nabla_{V_j}\big(G(\mu_i)^\top
      [S^{\frac{1}{2}}(\mu_{i}) \Phi  S^{\frac{1}{2}}(\mu_{i})]^{-1} \big) (\mu_i - X_i) \\
      + G(\mu_i)^\top [S^{\frac{1}{2}}(\mu_{i}) \Phi  S^{\frac{1}{2}}(\mu_{i})]^{-1} \nabla_{V_j} \big( \mu_i - X_i\big)\bigg)\Lambda_i\bigg]\\
      = \sum_{j=1}^p G(\mu_{ij})^2 [S^{\frac{1}{2}}(\mu_{ij}) \Phi_{jj}  S^{\frac{1}{2}}(\mu_{ij})]^{-1}  \frac{\partial \mu_{ij}}{\partial \eta_{ij}}
      \frac{\partial \eta_{ij}}{\partial V_j} \Lambda_i
      = \Lambda_i  \bigg(\text{Diag}(G(\mu_i)^2)[S^{\frac{1}{2}}(\mu_{i}) \Phi  S^{\frac{1}{2}}(\mu_{i})]^{-1} \bigg) \Lambda_i^\top
\end{multline}
where the first component vanishes because $\Exp(X_i - \mu_i) = \mathbf{0}_p$.
\end{description}